\documentclass[%
 reprint,
 amsmath,amssymb,
 aps,
pra,
]{revtex4-2}

\usepackage{graphicx}% Include figure files
\usepackage{dcolumn}% Align table columns on decimal point
\usepackage{bm}% bold math
\usepackage{physics}
\usepackage{hyperref}% add hypertext capabilities
\usepackage[caption=false]{subfig}
\usepackage{booktabs}
\usepackage{moreverb}
\usepackage{color}
\immediate\write18{texcount -inc -incbib 
-sum borra.tex > /tmp/wordcount.tex}

\immediate\write18{texcount -char -freq
 borra.tex > /tmp/charcount.tex}

\begin{document}

\preprint{arXiv}

\title{Quantum Information Propagation Preserving Computational Electromagnetics
}% Force line breaks with \\
\author{Dong-Yeop Na}
\affiliation{%
School of Electrical and Computer Engineering, Purdue University, West Lafayette IN, 47906, USA\\
}%

\author{Jie Zhu}%
\affiliation{%
School of Electrical and Computer Engineering, Purdue University, West Lafayette IN, 47906, USA\\
}%

\author{Fernando L. Teixeira}%
\affiliation{%
ElectroScience Laboratory and Department of Electrical and Computer Engineering, The Ohio State University, Columbus OH, 43212, USA\\
}%

\author{Weng C. Chew}
\email{wcchew@purdue.edu} 
\affiliation{%
School of Electrical and Computer Engineering, Purdue University, West Lafayette IN, 47906, USA\\
}%

\date{\today}% It is always \today, today,

\begin{abstract}
We propose a new methodology, called {\it numerical canonical quantization}, to solve quantum Maxwell's equations useful for mathematical modeling of quantum optics physics, and numerical experiments on arbitrary passive and lossless quantum-optical systems.  It is based on: (1) the macroscopic (phenomenological) electromagnetic theory on quantum electrodynamics (QED), and (2) concepts borrowed from computational electromagnetics.
It was shown that canonical quantization in inhomogeneous dielectric media required definite and proper normal modes (L. Kn\"oll, W. Vogel, and D.-G. Welsch, Phys. Rev. A {\bf 36}, 3803 (1987) and R. Glauber and M. Lewenstein, Phys. Rev. A {\bf 43}, 467 (1991)).
Here, instead of {\it ad-hoc} analytic normal modes, we numerically construct complete and time-reversible normal modes in the form of traveling waves to diagonalize the Hamiltonian.
Specifically, we directly solve the Helmholtz wave equations for a general linear, reciprocal, isotropic, non-dispersive, and inhomogeneous dielectric media by using either finite-element or finite-difference methods.
To convert a scattering problem with infinite number of modes into one with a finite number of modes, we impose {\it Bloch-periodic boundary conditions}.  This will sparsely sample the normal modes with numerical Bloch-Floquet-like normal modes.
Subsequent procedure of numerical canonical quantization is straightforward using linear algebra.
We provide relevant numerical recipes in detail and show an important numerical example of indistinguishable two-photon interference in quantum beam splitters, exhibiting Hong-Ou-Mandel effect, which is purely a quantum effect.
Also, the present methodology provides a way of numerically investigating existing or new macroscopic QED theories.  It will eventually allow quantum-optical numerical experiments of high fidelity to replace many real experiments as in classical electromagnetics.
As in classical electromagnetics, furthermore, this methodology would provide numerical experiments to explore a large variety of quantum-optical phenomena in complex systems that still perplex the community:  cases in point are virtual photons, quantum-optical metamaterials, superluminal photons by quantum tunneling, interaction free measurement, NOON states in quantum metrology and quantum sensing, and time entanglement.
\end{abstract}

%Use showkeys class option if keyword
\maketitle

%\tableofcontents

\section{Introduction}
The recent advent in quantum computing spells the beginning of an exciting era for quantum technologies.
Mathematical modeling of physical phenomena and their numerical simulations have transformed classical electromagnetics technologies.  But such knowledge base is still in its infancy for quantum Maxwell's equations and quantum technologies.
Quantum Maxwell's equations (QME) \cite{CHEW2016quantum,cohen1998atom,Mandel1995optical,SCHEEL2008numerical},
where classical Maxwell dynamic field and source variables are
elevated to infinite-dimensional quantum {\it operators} are
\begin{flalign}
\nabla \times \hat{\mathbf{E}}&=-\frac{\partial \hat{\mathbf{B}}}{\partial t},
\nonumber
\\
\nabla \times \hat{\mathbf{H}}&=\hat{\mathbf{J}}+\frac{\partial \hat{\mathbf{D}}}{\partial t},
\nonumber
\\
\nabla \cdot \hat{\mathbf{D}}&=\hat{\rho},
\nonumber
\\
\nabla \cdot \hat{\mathbf{B}}&=0.
 \label{eq:QME}
\end{flalign}
The above quantum equations are rigorously derived in the Heisenberg picture, in both
coordinate and mode Hilbert spaces, in \cite{CHEW2016quantum}, and extended to inhomogeneous media with impressed sources.  The above quantum operator equations are meaningful only when they operate on a quantum state. The corresponding equation of motion for the quantum state is
\begin{flalign}
\hat{H}\ket{\psi\left(t\right)}=i\hbar\frac{\partial}{\partial t}\ket{\psi\left(t\right)}
\label{eq:QSEa}
\end{flalign}
and the full description of the Hamiltonian is given in \cite{CHEW2016quantum}.
The Hamiltonian is easily diagonalizable in the mode Hilbert space making it mathematically homomorphic (analogous) to harmonic oscillators uncoupled from each other.  This is the popular approach in quantum optics (QO).
Then, using this QO approach, we characterize how photons, transported in the form of electromagnetic (EM) fields, interfere with each other.
Together with quantum theory, it has spurred the development of various quantum technologies \cite{Nielsen2011quantum,Gisin2002quantum,Lanzagorta2011quantum,Shih2007quantum,vogel2006quantum,fox2006quantum}
in which few-photon interference and their granularity are important for their quantum effects.

Nevertheless, early pioneers have mathematically shown that one can employ the {\it macroscopic} framework on quantum electrodynamics (QED) which admits the partial use of the macroscopic Maxwell's theory.  This theory uses the concept of effective electromagnetic media \cite{Jauch1948phenomenological,Glauber1991quantum,SCHEEL2008numerical}, when the optical/microwave photon wavelengths are typically much larger than the
size of the atoms or molecules.
That is, lossless dielectric
objects can be modeled by inhomogeneous permittivity over a bandwidth which dispersion effect is negligible.
Moreover, it is important to find the normal modes in the form of traveling waves for canonical quantization in these media \cite{Glauber1991quantum,Knoll1987action}, \footnote{In particular, Bloch-Floquet normal modes were used for canonical quantization in periodic structures for the study of characterizing spontaneous emission properties in photonic crystals \cite{Kweon1995quantum}.  But our interest in using Bloch-Floquet modes is quite different, viz., to convert a problem that has infinite number of modes into one with a finite number of modes.}.

More recently, quantization procedures in general dispersive, dissipative, anisotropic, or reciprocal media
\cite{Dalton1996field,Crenshaw2008comparison,Adler1962quantum,SCHEEL2008numerical,Dung1998three,Philbin2010canonical,CHEW2016quantum,Sha2018dissipative}
have been developed.
A variety of quantum-optical phenomena can be captured by macroscopic QED theory, such as spontaneous emissions
in dielectric media \cite{Crenshaw2008comparison}, Casimir forces
\cite{Philbin2011Casimir}, artificial atoms \cite{Bratschitsch2006artificial}, few-photon interference in passive and lossless quantum-optical instruments \cite{Dumke2016roadmap}, quantum-optical metamaterials \cite{Georgi2019metasurface}, NOON states in quantum metrology and quantum sensing \cite{Zhang2018interfacing}, just to name a few.

Most of previous numerical works for few-photon interference in quantum-optical systems adopted the atomistic approach  where multi-level atoms interact via free fields in free space.
As such, infinite number of plane waves are used to expand the free fields.
%Note that one can do arbitrary sampling and truncation with the closed-form normal modes.
To reduce the complexity of the problem, cavity QED \cite{BUZEK1999numerical} has been used to analyze QED effects on spontaneous emission of an excited two-level atom inside a cavity.
And \cite{HAVUKAINEN1999quantum,LONGO2009dynamics,LONGO2012hong}
represented free fields by a set of plane waves to investigate the quantum beam splitter (QBS), which is modeled by multi-level atoms.
Despite its interesting physics, the atomistic model is not suited for dealing with complex macroscopic problems  involving a large number of atoms.  Also, for complex problems, normal modes do not have closed-form solutions.

%Our recent efforts \cite{CHEW2016quantum}
%seek to build bridges between computational electromagnetics (CEM) and physics
%communities to foster cross pollination.
%This work is along the same spirit of \cite{CHEW2016quantum} but with focus on some concrete examples, aimed at %demonstrating the strength of numerical implementation of macroscopic QED theory with its germane physical phenomena.

\begin{figure}
\centering
\includegraphics[width=\linewidth]
{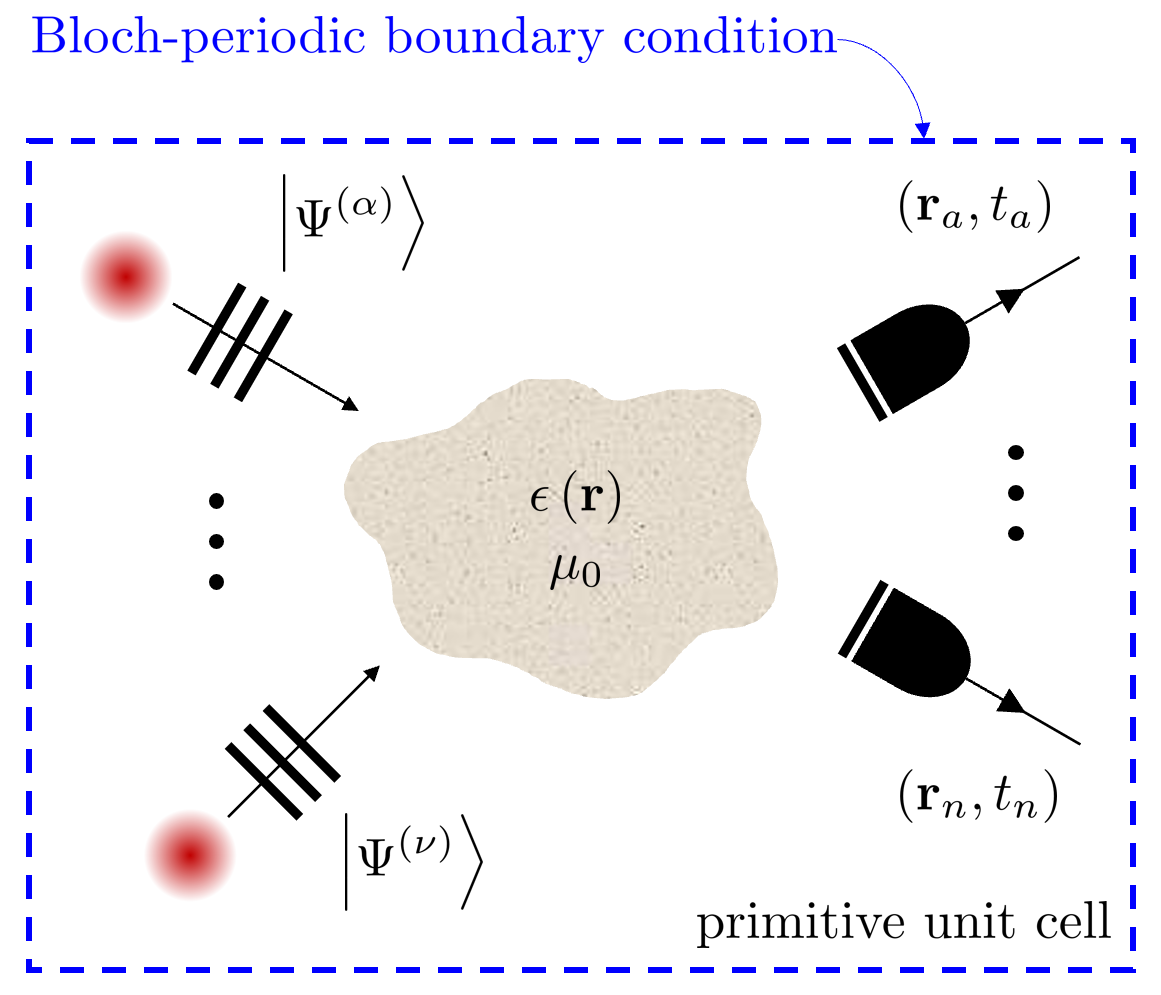}
\caption{Illustration of modeling arbitrary passive and lossless quantum optical systems driven by few photons.}
\label{fig:illustration}
\end{figure}

In this paper, we propose a {\it numerical canonical quantization methodology}, useful for numerical experiments on arbitrary passive and lossless quantum-optical systems for non-dispersive media, as illustrated in Fig. \ref{fig:illustration}.  This is valid over the bandwidth where frequency dispersive effect can be ignored.
Based on the macroscopic QED theory, such systems can be modeled by linear, reciprocal, isotropic, non-dispersive, and inhomogeneous dielectric media.
But here, instead of {\it ad-hoc} analytic normal modes, we construct complete and time-reversible numerical normal modes in the form of traveling waves to diagonalize the Hamiltonian.
In other words, borrowing the concept of computational electromagnetics (CEM), we directly solve the Helmholtz wave equations for the media by using either finite-element or finite-difference methods on a given mesh.

%For appropriate boundary conditions, we further assume that the system is in a primitive unit cell, implying that the resulting media are periodic in which {\it Bloch-Floquet} normal modes form a complete set of orthogonal functions.
%The boundary conditions correspond to {\it Bloch-periodic boundary condition} (B-PBC).
%However, it should be emphasized that the physics of interest is {\it local dynamics} of photons interacting with dielectric bodies inside the primitive unit cell.% but not the global quantum physics over whole photonic crystals.
%This argument stands on the interpretation for photons to be spatially and temporally localized in the form of wave %packets.

%Thus, the time evolution of the primitive local quantities should be isolated from any aliasing or interference of neighbor unit cells and observed within a properly-chosen time window.
%Since the primitive local fields consist of very high frequency components of normal modes, we model them through the linear superposition of numerical Bloch-Floquet-like normal modes sparsely sampled at integer multiples of the reciprocal lattice vector.
%Eventually, we solve a finite-dimensional generalized eigenvalue problem that returns the countably-finite number of the numerical normal modes so as to facilitate computation.
%Subsequent procedures of numerical canonical quantization are straightforward using linear algebra.
%We provide relevant numerical recipes in detail and show an important numerical example of indistinguishable two-photon interference in quantum beam splitters (QBSs), exhibiting Hong-Ou-Mandel effect.

The present methodology provides a way of using new macroscopic QED theory to mathematically model quantum-optical systems with little physical restrictions in real experimental set-up.
Eventually, as has happened in the classical electromagnetic case, it would provide a numerical experimental platform to explore a large variety of perplexing QED/quantum-optical phenomena such as virtual photons, quantum-optical metamaterials, superluminal photons by quantum tunneling, interaction free measurement, quantum metrology and quantum sensing, and time entanglement \cite{Gerry2013quantum}.  This is increasingly important as quantum technologies become increasingly complex.

The numerical canonical quantization here resembles the analytical canonical quantization.  The difference is that the modes are found numerically, and then quantized analytically.   Then quantum information can be injected into these modes displaying their quantum effects.  The modes are formed into wave packets, and we liken the photons to be ``riding" on the wave packet formed by these modes.

Note that the time convention $e^{-i\omega t}$ is suppressed
throughout this work.

\section{Numerical canonical quantization in periodic dielectric media}

In this section, we briefly revisit the canonical quantization for periodic dielectric media but via Bloch-Floquet normal modes \footnote{It is to be noted that Bloch-Floquet modes are not our main focus, but a mathematical trick to convert the uncountably-infinite modes to countably-infinite modes.}.

\subsection{Hamiltonian density and Hamilton's equations of motion}
Consider a linear, reciprocal, lossless, isotropic, non-dispersive, and periodic dielectric medium, namely, the permittivity can be written as
\begin{flalign}
\epsilon=\epsilon\left(\mathbf{r}\right)=\epsilon\left(\mathbf{r}+\mathbf{R}\right)
\end{flalign}
where $\mathbf{R}$ is a primitive lattice translation vector.
Vector potential $\mathbf{A}\left(\mathbf{r},t\right)$ should be associated with scalar potential $\Phi\left(\mathbf{r},t\right)$ through the general Lorenz gauge \cite{CHEW2014vector}.
Specifically, when external sources are placed infinitely far away (or absent) from a region of interest, it becomes equivalent to $\Phi=0$ or the generalized Coulomb gauge \cite{LIU2018potential}.
Thus,
\begin{flalign}
\nabla\cdot\epsilon\left(\mathbf{r}\right)\mathbf{A}\left(\mathbf{r},t\right)
&=0
\label{eq:tr_gauge}
\end{flalign}
and the classical Hamiltonian density $\mathcal{H}\left(\mathbf{r},t\right)$, involved in the vector potential only, can be defined \cite{CHEW2016quantum} by
\begin{flalign}
\mathcal{H}\left(\mathbf{r},t\right)
&=
\frac{1}{2\epsilon\left(\mathbf{r}\right)}
\dot{\mathbf{A}}\left(\mathbf{r},t\right)
\cdot
\dot{\mathbf{A}}\left(\mathbf{r},t\right)
\nonumber\\
&+
\frac{1}{2\mu_{0}}
\nabla\times\mathbf{A}\left(\mathbf{r},t\right)
\cdot
\nabla\times\mathbf{A}\left(\mathbf{r},t\right)
\label{eq:Hamiltonian_density}
\end{flalign}
where $\dot{\mathbf{A}}\left(\mathbf{r},t\right)=\partial \mathbf{A}\left(\mathbf{r},t\right)/\partial t$.
The total Hamiltonian $H$ is given by
\begin{align}
H=\int d\mathbf{r} \mathcal{H}\left(\mathbf{r},t\right).
\label{eq:Hamiltonian_0}
\end{align}
Note that, because of energy conservation, this Hamiltonian should be time
independent.
Then, the classical Hamilton's equations of motion, which are vector wave equations, can be written by
\begin{flalign}
\nabla\times\nabla\times
\mathbf{A}\left(\mathbf{r},t\right)
+
\epsilon\left(\mathbf{r}\right)\mu_{0}
\ddot{\mathbf{A}}\left(\mathbf{r},t\right)
=
0.
\label{eq:wave_eqn_vec_pot_1}
\end{flalign}
Furthermore, for the transverse electric (TE) polarization of vector potentials in 2-D problems, since $\nabla\cdot\mathbf{A}\left(\mathbf{r}\right)=0$, (\ref{eq:wave_eqn_vec_pot_1}) reduces to
\begin{flalign}
\nabla^{2}
\mathbf{A}\left(\mathbf{r},t\right)
-
\epsilon\left(\mathbf{r}\right)\mu_{0}
\ddot{\mathbf{A}}\left(\mathbf{r},t\right)
=
0
\label{eq:wave_eqn_vec_pot_2}
\end{flalign}
that does not include the null space.  In this work, these modes are found numerically.

\subsection{Vectorial mode decomposition via Bloch-Floquet modes}
Due to the linearity of media and realness of vector potentials, one can represent a time-harmonic or monochromatic vector potential using the separation of variables (SOV) with respect to space and time as
\begin{flalign}
\mathbf{A}_{\omega}\left(\mathbf{r},t\right)
=
\alpha_{\omega}\left(t\right)
\tilde{\mathbf{A}}_{\omega}\left(\mathbf{r}\right)
=
\alpha^{*}_{\omega}\left(t\right)
\tilde{\mathbf{A}}^{*}_{\omega}\left(\mathbf{r}\right).
\label{eq:vec_pot_3D_sov_rep_1}
\end{flalign}
Then, substituting (\ref{eq:vec_pot_3D_sov_rep_1}) into (\ref{eq:wave_eqn_vec_pot_2}) yields two decoupled differential equations linked by a separation constant $-\omega^{2}$; viz.,
\begin{flalign}
\ddot{\alpha}_{\omega}\left(t\right)
&=-\omega^{2} \alpha_{\omega}\left(t\right)
\\
\nabla^{2}\tilde{\mathbf{A}}_{\omega}\left(\mathbf{r}\right)
&=
-\epsilon\left(\mathbf{r}\right)\mu_{0}
\omega^{2}
\tilde{\mathbf{A}}_{\omega}\left(\mathbf{r}\right).
\label{eq:Helmholtz_eq}
\end{flalign}
General solutions of $\alpha_{\omega}\left(t\right)$ are given by the linear combination of $e^{\mp i\omega t}$.
A vectorial eigenfunction of the generalized eigenvalue problem (\ref{eq:Helmholtz_eq}) is known as {\it Bloch-Floquet} mode that takes the form of
\begin{flalign}
\tilde{\mathbf{A}}_{\omega}\left(\mathbf{r}\right)=\mathbf{u}_{\boldsymbol{\kappa}}\left(\mathbf{r}\right)e^{i\boldsymbol{\kappa}\cdot\mathbf{r}}
\end{flalign}
where $\boldsymbol{\kappa}$ is Bloch wavevector and $\mathbf{u}_{\boldsymbol{\kappa}}\left(\mathbf{r}\right)$ is a periodic function by the primitive translation vector.
It is important to note that, in most cases, Bloch-Floquet modes are {\it time-reversible} \footnote{The time-reversibility is closely related to the medium reciprocity. For example, it is broken in gyrotropic or magneto-optical media which are described by $3\times 3$ complex Hermitian dielectric tensors (lossless) but are non-reciprocal \cite{CHEW1996waves,Joannopoulos2008photonic}.}.
This fact induces two important properties: (1) Two degenerate vectorial eigenfunctions, having $\pm\boldsymbol{\kappa}$, are supposed to share the same eigenfrequency, i.e., $\omega_{\boldsymbol{\kappa}}=\omega_{-\boldsymbol{\kappa}}$; and (2) they are related by complex conjugate as
\begin{flalign}
\tilde{\mathbf{A}}_{-\boldsymbol{\kappa}}\left(\mathbf{r}\right)
=
\tilde{\mathbf{A}}_{\boldsymbol{\kappa}}^{*}\left(\mathbf{r}\right).
\label{eq:time_reversal_invariance}
\end{flalign}
Hence, a general monochromatic solution $\tilde{\mathbf{A}}_{\omega}\left(\mathbf{r}\right)$ can be given by the degenerate pair's linear superposition.
Eventually, arbitrary real-valued vector potentials in periodic dielectric media can be expanded by Bloch-Floquet normal modes over $\boldsymbol{\kappa}$
\begin{flalign}
\mathbf{A}\left(\mathbf{r},t\right)
=
\sum_{\boldsymbol{\kappa}}
q_{\boldsymbol{\kappa}}\left(t\right)
\tilde{\mathbf{A}}_{\boldsymbol{\kappa}}\left(\mathbf{r}\right)
+
q^{*}_{\boldsymbol{\kappa}}\left(t\right)
\tilde{\mathbf{A}}_{\boldsymbol{\kappa}}^{*}\left(\mathbf{r}\right)
\label{eq:vec_pot_3D_sov_rep_2}
\end{flalign}
where $q_{\boldsymbol{\kappa}}\left(t\right)=q_{\boldsymbol{\kappa}}\left(t=0\right)e^{-i\omega_{\boldsymbol{\kappa}}t}$.
Notice that $\tilde{\mathbf{A}}_{\boldsymbol{\kappa}}\left(\mathbf{r}\right)$ obtained from (\ref{eq:Helmholtz_eq}) should satisfy the following orthonormal condition \cite{CHEW2016quantum}
\begin{flalign}
\mel
{\tilde{\mathbf{A}}_{\boldsymbol{\kappa}}}
{\epsilon}
{\tilde{\mathbf{A}}_{\boldsymbol{\kappa}'}}
&\equiv
\iiint \limits_{\Omega}d\mathbf{r}
\left(
\epsilon\left(\mathbf{r}\right)
\tilde{\mathbf{A}}^{*}_{\boldsymbol{\kappa}}\left(\mathbf{r}\right)
\cdot
\tilde{\mathbf{A}}_{\boldsymbol{\kappa}'}\left(\mathbf{r}\right)
\right)
\nonumber\\
&=\delta_{\boldsymbol{\kappa},\boldsymbol{\kappa}'}
\label{eq:orthonormality}
\end{flalign}
where $\Omega$ denotes the primitive unit cell volume and $\delta$ is the Kronecker delta function.

\begin{figure*}
\centering
\subfloat[Single input of classical light beam]{\includegraphics[width=.25\linewidth]{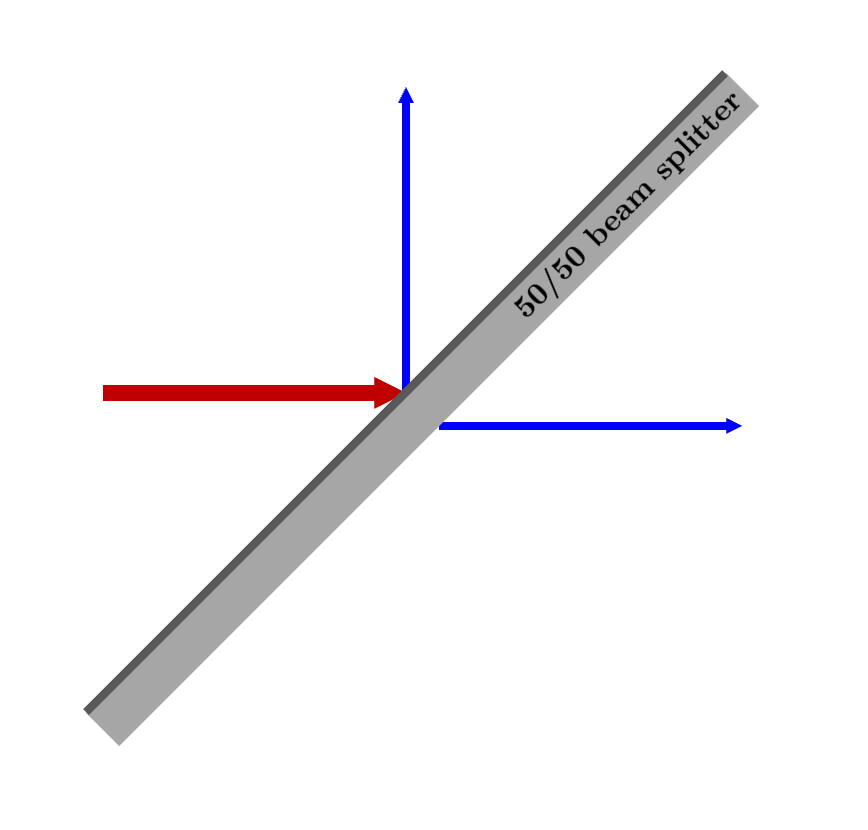}
\label{fig:CBS_schm_1}
}
\subfloat[Both inputs of classical light beams]{\includegraphics[width=.25\linewidth]{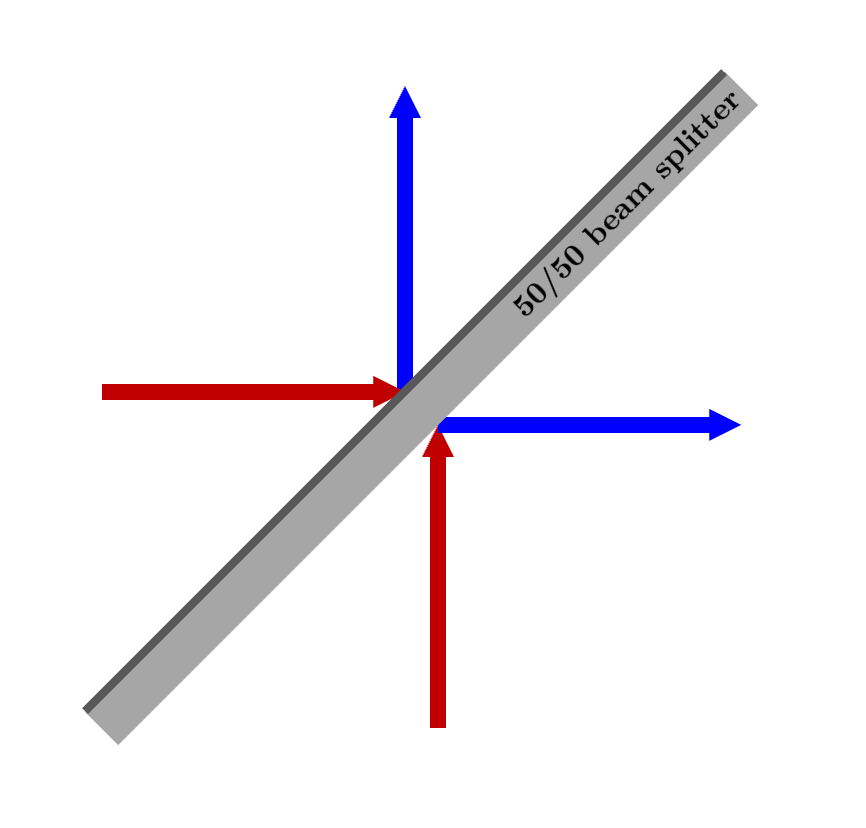}
\label{fig:CBS_schm_2}
}
\subfloat[Single-photon incidence]{\includegraphics[width=.25\linewidth]{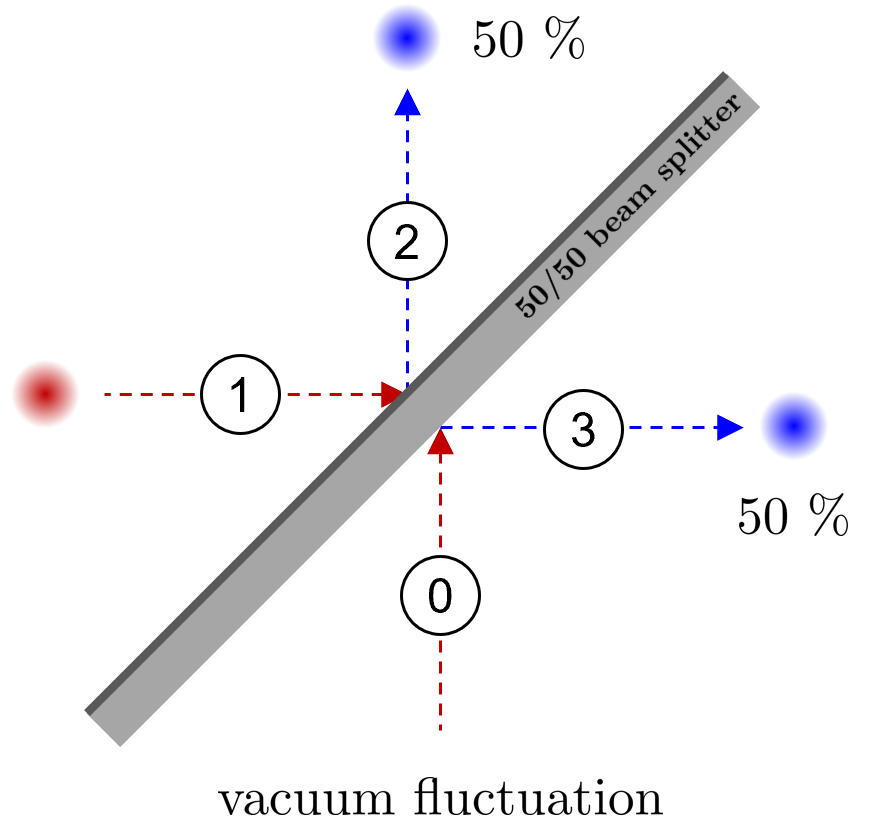}
\label{fig:QBS_schm_1}
}
\subfloat[Two-photon incidence]{\includegraphics[width=.25\linewidth]{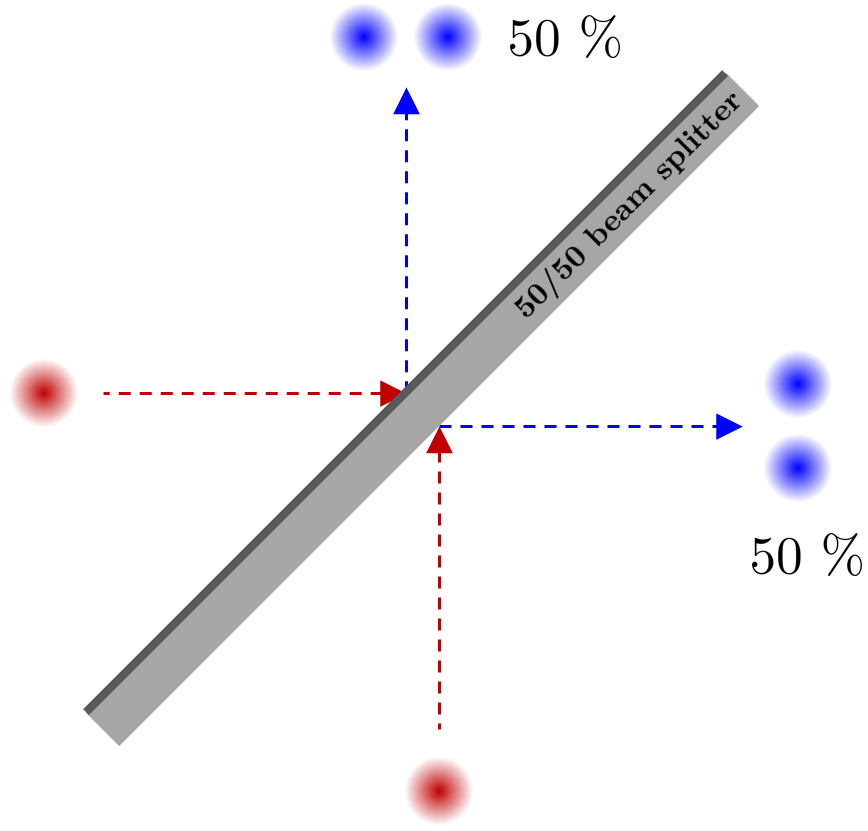}
\label{fig:QBS_schm_2}
}
\caption{Schematics for classical and quantum beam splitters for various input scenarios.}
\label{fig:QBS_cases}
\end{figure*}

\subsection{Diagonalization of Hamiltonian and canonical quantization}
Substituting (\ref{eq:vec_pot_3D_sov_rep_2}) into (\ref{eq:Hamiltonian_density}), taking a volume integral over $\Omega$, and utilizing integral by parts and the orthonormal relation (\ref{eq:orthonormality}), one can rewrite the Hamiltonian as
\begin{flalign}
H
&=
\sum_{\boldsymbol{\kappa},\boldsymbol{\kappa}'}
\Bigl(
\dot{q}_{\boldsymbol{\kappa}}\left(t\right)
\dot{q}_{\boldsymbol{\kappa}'}\left(t\right)
\delta_{-\boldsymbol{\kappa},\boldsymbol{\kappa}'}
+
\dot{q}_{\boldsymbol{\kappa}}\left(t\right)
\dot{q}_{\boldsymbol{\kappa}'}^{*}\left(t\right)
\delta_{\boldsymbol{\kappa},\boldsymbol{\kappa}'}
\nonumber \\
&+
\dot{q}_{\boldsymbol{\kappa}}^{*}\left(t\right)
\dot{q}_{\boldsymbol{\kappa}'}\left(t\right)
\delta_{\boldsymbol{\kappa},\boldsymbol{\kappa}'}
+
\dot{q}_{\boldsymbol{\kappa}}^{*}\left(t\right)
\dot{q}_{\boldsymbol{\kappa}'}^{*}\left(t\right)
\delta_{-\boldsymbol{\kappa},\boldsymbol{\kappa}'}
\nonumber \\
&+
\omega_{\boldsymbol{\kappa}}^{2}q_{\boldsymbol{\kappa}}\left(t\right)
q_{\boldsymbol{\kappa}'}\left(t\right)
\delta_{-\boldsymbol{\kappa},\boldsymbol{\kappa}'}
+
\omega_{\boldsymbol{\kappa}}^{2}q_{\boldsymbol{\kappa}}\left(t\right)
q_{\boldsymbol{\kappa}'}^{*}\left(t\right)
\delta_{\boldsymbol{\kappa},\boldsymbol{\kappa}'}
\nonumber \\
&+
\omega_{\boldsymbol{\kappa}}^{2}q_{\boldsymbol{\kappa}}^{*}\left(t\right)
q_{\boldsymbol{\kappa}'}\left(t\right)
\delta_{\boldsymbol{\kappa},\boldsymbol{\kappa}'}
+
\omega_{\boldsymbol{\kappa}}^{2}q_{\boldsymbol{\kappa}}^{*}\left(t\right)
q_{\boldsymbol{\kappa}'}^{*}\left(t\right)
\delta_{-\boldsymbol{\kappa},\boldsymbol{\kappa}'}
\Bigr).
\end{flalign}
It should be mentioned that $\delta_{-\boldsymbol{\kappa},\boldsymbol{\kappa}'}$ results from the orthonormal relation (\ref{eq:orthonormality}) and time-reversibility (\ref{eq:time_reversal_invariance}).
Since $\dot{q}_{\boldsymbol{\kappa}}\left(t\right)=-i\omega_{\boldsymbol{\kappa}}q_{\boldsymbol{\kappa}}\left(t\right)$, all anti-diagonal terms multiplied by $\delta_{-\boldsymbol{\kappa},\boldsymbol{\kappa}'}$ are canceled out, as a result, the Hamiltonian is summed over a single dummy index $\boldsymbol{\kappa}$ as
\begin{flalign}
H
&=
\sum_{\boldsymbol{\kappa}}
\dot{q}_{\boldsymbol{\kappa}}\left(t\right)\dot{q}_{\boldsymbol{\kappa}'}^{*}\left(t\right)
+
\omega_{\boldsymbol{\kappa}}^{2}
q_{\boldsymbol{\kappa}}\left(t\right)q_{\boldsymbol{\kappa}'}^{*}\left(t\right)
\nonumber\\
&=
\sum_{\boldsymbol{\kappa}}
\left|P_{\boldsymbol{\kappa}}\left(t\right)\right|^{2}
+
\left|Q_{\boldsymbol{\kappa}}\left(t\right)\right|^{2}
=
\sum_{\boldsymbol{\kappa}}
B_{\boldsymbol{\kappa}}\left(t\right)
B^{*}_{\boldsymbol{\kappa}}\left(t\right)
\label{eq:Hamiltonian_3}
\end{flalign}
where conjugate canonical variables $P_{\boldsymbol{\kappa}}\left(t\right)=\dot{q}_{\boldsymbol{\kappa}}\left(t\right)$, $Q_{\boldsymbol{\kappa}}\left(t\right)=\omega_{\boldsymbol{\kappa}}q_{\boldsymbol{\kappa}}\left(t\right)$, and $B_{\boldsymbol{\kappa}}\left(t\right)=i P_{\boldsymbol{\kappa}}\left(t\right)+Q_{\boldsymbol{\kappa}}\left(t\right)$.
Note that the above reduction from double sum to a single sum is the hallmark of the diagonalization of the Hamiltonian.
It can be seen that the resulting Hamiltonian (\ref{eq:Hamiltonian_3}) is mathematically homomorphic to the sum of Hamiltonians of many lone harmonic oscillators or classical pendulums.

%%%%%%%%%%%%%%%%%%%%%%%%%%%%%%%%%%%%%%%%%%%%%%%%%%%%%%%%%%%%%%%%%%%
At this juncture, we can introduce dimensionless amplitude
${a}_{\boldsymbol{\kappa}}\left(t\right)$ such that
\begin{align}
B_{\boldsymbol{\kappa}}\left(t\right)=\sqrt{\hbar\omega_{\boldsymbol{\kappa}}}{a}_{\boldsymbol{\kappa}}\left(t\right)
\end{align}
The classical Hamiltonian then becomes
\begin{flalign}
{H}
=
\sum_{\boldsymbol{\kappa}}
\hbar\omega_{\boldsymbol{\kappa}}
{a}^{*}_{\boldsymbol{\kappa}}\left(t\right) {a}_{\boldsymbol{\kappa}}\left(t\right)
=
\sum_{\boldsymbol{\kappa}}
{H}_{\boldsymbol{\kappa}}.
\label{eq:q_hamiltonian}
\end{flalign}
The above is just the classical Hamiltonian for a sum of harmonic oscillators.
This is still a classical Hamiltonian, but we can now define our conjugate variables as
${a}^{*}_{\boldsymbol{\kappa}}\left(t\right)$ and ${a}_{\boldsymbol{\kappa}}\left(t\right)$.  The equations of motion (EOM) for
${a}^{*}_{\boldsymbol{\kappa}}\left(t\right)$ and ${a}_{\boldsymbol{\kappa}}\left(t\right)$ can be derived using energy conservation argument to give \cite{Chew2019Hamiltonian} \footnote{We pick the time-stretching constant so the EOM's agree with known EOM's.}
\begin{align}
\frac{d a_{\boldsymbol{\kappa}}(t)}{dt}=-\frac{i}{\hbar}\frac{\partial H}{\partial a^*_{\boldsymbol{\kappa}}(t)},\quad
\frac{d a^*_{\boldsymbol{\kappa}}(t)}{dt}=\frac{i}{\hbar}\frac{\partial H}{\partial a_{\boldsymbol{\kappa}}(t)}.
\end{align}
%*********************************************************************************************************
The above yields
%
%\begin{align}
%\frac{d a_{\boldsymbol{\kappa}}(t)}{dt}=-i\hbar \frac{\partial H}{\partial a^*_{\boldsymbol{\kappa}(t)}},\quad
%\frac{d a^*_{\boldsymbol{\kappa}}(t)}{dt}=-i\frac{\partial H}{\partial a_{\boldsymbol{\kappa}}(t)},
%\end{align}
\begin{align}
\frac{d a_{\boldsymbol{\kappa}}(t)}{dt}=-i\omega_{\boldsymbol{\kappa}}  a_{\boldsymbol{\kappa}}(t),\quad
\frac{d a^*_{\boldsymbol{\kappa}}(t)}{dt}=i\omega_{\boldsymbol{\kappa}}  a^*_{\boldsymbol{\kappa}}(t).
\end{align}
The above are the EOM of a classical pendulum expressed in rotating wave picture.

Now to turn a classical Hamiltonian $H$ into a quantum Hamiltonian $\hat H$, we elevate the conjugate variables
${a}^{*}_{\boldsymbol{\kappa}}\left(t\right)$ and ${a}_{\boldsymbol{\kappa}}\left(t\right)$ to become quantum operators
$\hat{a}^{\dag}_{\boldsymbol{\kappa}}\left(t\right)$ and $\hat{a}_{\boldsymbol{\kappa}}\left(t\right)$ which are the quantum creation and annihilation operators for the quantum harmonic oscillator.

%
%This makes it easy to connect them to the lone quantum harmonic oscillators; thus,
% one can obtain quantum Hamiltonian operators $\hat{H}$ from (\ref{eq:Hamiltonian_3}) by the following inference:
%\begin{flalign}
%B_{\boldsymbol{\kappa}}\left(t\right)&\rightarrow\sqrt{2\hbar\omega_{\boldsymbol{\kappa}}}\hat{a}_{\boldsymbol{\kappa}}\left(t\right),
%\\
%B^{*}_{\boldsymbol{\kappa}}\left(t\right)&\rightarrow\sqrt{2\hbar\omega_{\boldsymbol{\kappa}}}\hat{a}_{\boldsymbol{\kappa}}^{\dag}\left(t\right),
%\end{flalign}
%where $\hat{a}$ and $\hat{a}^{\dag}$ are annihilation and creation operators, respectively, whose commutator relation is given by
%\begin{flalign}
%\left[\hat{a}_{\boldsymbol{\kappa}}\left(t\right),\hat{a}^{\dag}_{\boldsymbol{\kappa}'}\left(t\right)\right]
%&=\hat{a}_{\boldsymbol{\kappa}}\left(t\right)\hat{a}^{\dag}_{\boldsymbol{\kappa}'}\left(t\right)
%-\hat{a}^{\dag}_{\boldsymbol{\kappa}'}\left(t\right)\hat{a}_{\boldsymbol{\kappa}}\left(t\right)
%\nonumber \\
%&=\delta_{\boldsymbol{\kappa},\boldsymbol{\kappa}'}\hat{I},
%\\
%\left[\hat{a}_{\boldsymbol{\kappa}}\left(t\right),\hat{a}_{\boldsymbol{\kappa}'}\left(t\right)\right]
%&=
%\left[\hat{a}^{\dag}_{\boldsymbol{\kappa}}\left(t\right),\hat{a}^{\dag}_{\boldsymbol{\kappa}'}\left(t\right)\right]=0,
%\end{flalign}
%where $\hat{I}$ is an identity operator.
%In other words, a value is changed by an operator inherently involved in random variable process.

Consequently, the quantum Hamiltonian for this system is
\begin{flalign}
\hat{H}
&=
\frac{1}{2}
\sum_{\boldsymbol{\kappa}}
\hbar\omega_{\boldsymbol{\kappa}}
\left(
\hat{a}^{\dag}_{\boldsymbol{\kappa}}\left(t\right)\hat{a}_{\boldsymbol{\kappa}}\left(t\right)
+
\hat{a}_{\boldsymbol{\kappa}}\left(t\right)\hat{a}^{\dag}_{\boldsymbol{\kappa}}\left(t\right)
\right)
\nonumber \\
&=
\sum_{\boldsymbol{\kappa}}
\hbar\omega_{\boldsymbol{\kappa}}
\left(
\hat{a}^{\dag}_{\boldsymbol{\kappa}}\left(t\right)\hat{a}_{\boldsymbol{\kappa}}
\left(t\right)
+\frac{1}{2}
\right)
=
\sum_{\boldsymbol{\kappa}}
\hat{H}_{\boldsymbol{\kappa}}.
\label{eq:q_hamiltonian}
\end{flalign}

%Classical modal amplitudes are also elevated into ladder operators as
%\begin{flalign}
%q_{\boldsymbol{\kappa}}\left(t\right) &\rightarrow \sqrt{\frac{\hbar}{2\omega_{\boldsymbol{\kappa}}}}\hat{a}_{\boldsymbol{\kappa}}\left(t\right)
%=
%\sqrt{\frac{\hbar}{2\omega_{\boldsymbol{\kappa}}}}\hat{a}_{\boldsymbol{\kappa}}e^{-i\omega_{\boldsymbol{\kappa}}t},
%\\
%q^{*}_{\boldsymbol{\kappa}}\left(t\right) &\rightarrow \sqrt{\frac{\hbar}{2\omega_{\boldsymbol{\kappa}}}}\hat{a}_{\boldsymbol{\kappa}}^{\dag}\left(t\right)=
%\sqrt{\frac{\hbar}{2\omega_{\boldsymbol{\kappa}}}}\hat{a}_{\boldsymbol{\kappa}}^{\dag}
%e^{-i\omega_{\boldsymbol{\kappa}}t},
%\end{flalign}
Finally one can finally obtain quantum vector potential operators
\begin{flalign}
\hat{\mathbf{A}}\left(\mathbf{r},t\right)
=
\hat{\mathbf{A}}^{(+)}\left(\mathbf{r},t\right)
+
\hat{\mathbf{A}}^{(-)}\left(\mathbf{r},t\right)
\end{flalign}
where
\begin{flalign}
\hat{\mathbf{A}}^{(+)}\left(\mathbf{r},t\right)
&=
\sum_{\boldsymbol{\kappa}}
\sqrt{\frac{\hbar}{2\omega_{\boldsymbol{\kappa}}}}
\tilde{\mathbf{A}}_{\boldsymbol{\kappa}}\left(\mathbf{r}\right)e^{-i\omega_{\boldsymbol{\kappa}}t}\hat{a}_{\boldsymbol{\kappa}},
\\
\hat{\mathbf{A}}^{(-)}\left(\mathbf{r},t\right)
&=
\sum_{\boldsymbol{\kappa}}
\sqrt{\frac{\hbar}{2\omega_{\boldsymbol{\kappa}}}}
\tilde{\mathbf{A}}^{*}_{\boldsymbol{\kappa}}\left(\mathbf{r}\right)e^{i\omega_{\boldsymbol{\kappa}}t}\hat{a}^{\dag}_{\boldsymbol{\kappa}},
\label{eq:vec_final_expr}
\end{flalign}
which are often called positive and negative frequency components
\cite{hahn1996hilbert}. 
Note that one can check the physical dimension agreement by substituting (\ref{eq:vec_final_expr}) into an energy density operator with the fact that eigenfunctions implicitly include a factor $1/\sqrt{\Omega\epsilon\left(\mathbf{r}\right)}$ to meet the orthonormal condition (\ref{eq:orthonormality}).
In other words,
\begin{align}
z^{(+)}(t)=\frac1 {2\pi}\int_0^\infty Z(\omega) e^{-i\omega t} d\omega
\end{align}
where $Z(\omega)$ is the Fourier transform of $z(t)$ and similar
definition applies to $z^{(-)}(t)$.
Note that the positive and negative components have their own important quantum physical meaning, which will be addressed in Sec. \ref{sec:numerical_example} more in detail.

%The classical EM phenomena are characterized by classical Maxwell's equations that stipulate the time evolution of EM fields, consisting of scalar numbers, throughout space.
%In contrast, due to the inherent uncertainty, the quantum optical systems should be described by using two mathematical tools: (1) operators, i.e., observables, and (2) state vectors.
%The former corresponds to, for example, the vector potential operators derived in the mode Hilbert space which are solutions of QME while satisfying well defined the commutator relations.
%On the other hand, quantum state vectors, denoted by $\ket{\psi}$, describe specific states of quantum systems and are to be taken as inputs of operators.
%Note that this concept is also extensively used in control systems.
%As seen in (\ref{eq:q_hamiltonian}), the quantum Hamiltonian operators are analogous to those of an ensemble of independent quantum harmonic oscillators.
%Therefore, quantum state vectors can be further identified by seeking eigenstates of a quantum Hamiltonian operator for each mode in (\ref{eq:q_hamiltonian}).

Since an operator (i.e., observable) has meaning only when the operator is acting on a quantum state vector that describes the state of systems of interest, one needs to prepare a proper quantum state vector that is a solution of {\it quantum state equation} (QSE).
First proposed by Schr\"{o}dinger, QSE is the fundamental postulate of the quantum theory \cite{Chew2019Hamiltonian}, while describing the time evolution of quantum state vectors, and takes the form of
\begin{flalign}
\hat{H}\ket{\psi\left(t\right)}=i\hbar\frac{\partial}{\partial t}\ket{\psi\left(t\right)}
\label{eq:QSE}
\end{flalign}
where $\ket{\psi\left(t\right)}$ is a quantum state vector which is a function of time.
The quantum state vector can be decomposed by time-dependent and -independent parts as
\begin{flalign}
\ket{\psi\left(t\right)}
&=
\hat{U}\left(t-t_{0}\right)\ket{\psi\left(t=t_{0}\right)}
\nonumber\\
&=
e^{-i\frac{\hat{H}}{\hbar}\left(t-t_{0}\right)}\ket{\psi\left(t=t_{0}\right)}
\label{eq:QSV_time_ev}
\end{flalign}
where $t_{0}$ is an arbitrary time instant and $\hat{U}\left(t-t_{0}\right)$ is a unitary operator.
Note that (\ref{eq:QSV_time_ev}) always satisfies (\ref{eq:QSE}), which can be checked by back-substitution to (\ref{eq:QSE}), and the unitary operator is an exponential function taking the quantum Hamiltonian operator argument.
This implies that if one can find a complete set of eigenstates of $\hat{H}$ at an arbitrary time instant $t_{0}$ to expand $\ket{\psi\left(t=t_{0}\right)}$, the time evolution of quantum state vectors can be completely tracked by (\ref{eq:QSV_time_ev}).
The resulting eigenstates of the quantum Hamiltonian operator for a single mode are {\it photon number states} as
\begin{flalign}
\hat{H}_{\boldsymbol{\kappa}_{p}}\ket{n_{\boldsymbol{\kappa}_{p}}}
=
\hbar\omega_{\boldsymbol{\kappa}_{p}}\left(n_{\boldsymbol{\kappa}_{p}}+\frac{1}{2}\right)\ket{n_{\boldsymbol{\kappa}_{p}}}
=
E_{\boldsymbol{\kappa}_{p}}\ket{n_{\boldsymbol{\kappa}_{p}}}.
\end{flalign}
Because the quantum Hamiltonian operators are mutually commuting with respect to different mode index $\boldsymbol{\kappa}$, the resulting composite eigenstate is given by the tensor product of the photon number states over all possible modes, known as {\it multimode Fock states}.
Without loss of generality, we determine the quantum state vector at $t_{0}=0$ as
\begin{flalign}
\ket{\psi\left(t=0\right)}=\ket{n_{\boldsymbol{\kappa}_{1}}}\otimes \cdots \otimes \ket{n_{\boldsymbol{\kappa}_{p}}} \otimes \cdots
\end{flalign}
Then,
\begin{flalign}
\hat{H}\ket{\psi\left(t=0\right)}
&=
\sum_{p}
\hat{H}_{\boldsymbol{\kappa}_{p}}
\ket{n_{\boldsymbol{\kappa}_{1}}}\otimes \cdots \otimes \ket{n_{\boldsymbol{\kappa}_{p}}} \otimes \cdots
\nonumber \\
&=
\sum_{p}
E_{\boldsymbol{\kappa}_{p}}\ket{\psi\left(t=0\right)}
=
E\ket{\psi\left(t=0\right)}
\end{flalign}
where
\begin{flalign}
E=\sum_{p}\hbar\omega_{\boldsymbol{\kappa}_{p}}\left(n_{\boldsymbol{\kappa}_{p}}+\frac{1}{2}\right).
\end{flalign}
Therefore, the quantum Hamiltonian operator can be diagonalized once again via photon number states.
And, by (\ref{eq:QSV_time_ev}),
\begin{flalign}
\ket{\psi\left(t\right)}
&=
e^{-i\frac{\hat{H}}{\hbar}t}\ket{\psi\left(t=0\right)}
\nonumber \\
&=
e^{-i
\left(
\frac{\sum_{p'}\hbar\omega_{\boldsymbol{\kappa}_{p'}}\left(n_{\boldsymbol{\kappa}_{p'}}+\frac{1}{2}\right)}
{\hbar}
\right)
t}
\ket{\psi\left(t=0\right)}
\nonumber \\
&=
e^{-i
\left(
\frac{\sum_{p'}E_{\boldsymbol{\kappa}_{p}}}{\hbar}
\right)
t}
\ket{\psi\left(t=0\right)}
\nonumber \\
&=
e^{-i\frac{E}{\hbar}t}
\ket{\psi\left(t=0\right)}.
\label{eq:exp_func_HO}
\end{flalign}
Note that one can prove the second equality in (\ref{eq:exp_func_HO}) using Baker–Campbell–Hausdorff (BCH) formula.
Thus, by substituting (\ref{eq:exp_func_HO}) into the RHS of (\ref{eq:QSE}), it is easy to show that
\begin{flalign}
i\hbar\frac{\partial}{\partial t}\ket{\psi\left(t\right)}
&=
i\hbar
\left(
-i
\frac{\sum_{p}\hbar\omega_{\boldsymbol{\kappa}_{p}}\left(n_{\boldsymbol{\kappa}_{p}}+\frac{1}{2}\right)}
{\hbar}
\right)
\nonumber \\
&\times
e^{-i
\left(
\frac{\sum_{p'}\hbar\omega_{\boldsymbol{\kappa}_{p'}}\left(n_{\boldsymbol{\kappa}_{p'}}+\frac{1}{2}\right)}
{\hbar}
\right)
t}
\ket{\psi\left(t=0\right)}
\nonumber \\
&=
E
e^{-i\frac{E}{\hbar}t}
\ket{\psi\left(t=0\right)}
\nonumber\\
&=
\hat{H}
e^{-i\frac{E}{\hbar}t}
\ket{\psi\left(t=0\right)}
=
\hat{H}\ket{\psi\left(t\right)}.
\end{flalign}
Due to the bosonic property, any composite eigenstates can be obtained from the ground state by applying some product of the creation operators, that is,
\begin{flalign}
\ket{n_{\boldsymbol{\kappa}_{1}}}
\otimes \cdots \otimes
\ket{n_{\boldsymbol{\kappa}_{p}}}
\otimes \cdots
=
\frac
{\left(\hat{a}^{\dag}_{\boldsymbol{\kappa}_{1}}\right)^{n_{1}}\cdots\left(\hat{a}^{\dag}_{\boldsymbol{\kappa}_{p}}\right)^{n_{p}}\cdots}
{\sqrt{\ket{n_{\boldsymbol{\kappa}_{1}}}!\cdots \ket{n_{\boldsymbol{\kappa}_{p}}}!\cdots}}
\ket{\left\{0\right\}}
\label{eq:multimode_Fock_state}
\end{flalign}
where $\ket{\left\{0\right\}}$ denotes the vacuum state.
Again, note that one can assemble an arbitrary initial quantum state vector, i.e., $\ket{\psi\left(t=0\right)}$, by the linear superposition of (\ref{eq:multimode_Fock_state}) which will be explained later in detail to construct a quantum state vector for a single photon riding on a wavepacket.

It should be emphasized that, based on both solutions of QME and QSE, i.e., operators and quantum state vectors, our mathematical model forms the axiomatic system, armed with self-consistency and completeness, so that the dynamics of arbitrary passive and lossless quantum optical systems can be properly captured.
Specifically, the second diagonalization of the quantum Hamiltonian operators by number states makes it possible to evaluate the expectation values of observables such as the energy density operator or arbitrary degrees of correlation functions.

\begin{figure}
\centering
{\includegraphics[width=\linewidth]{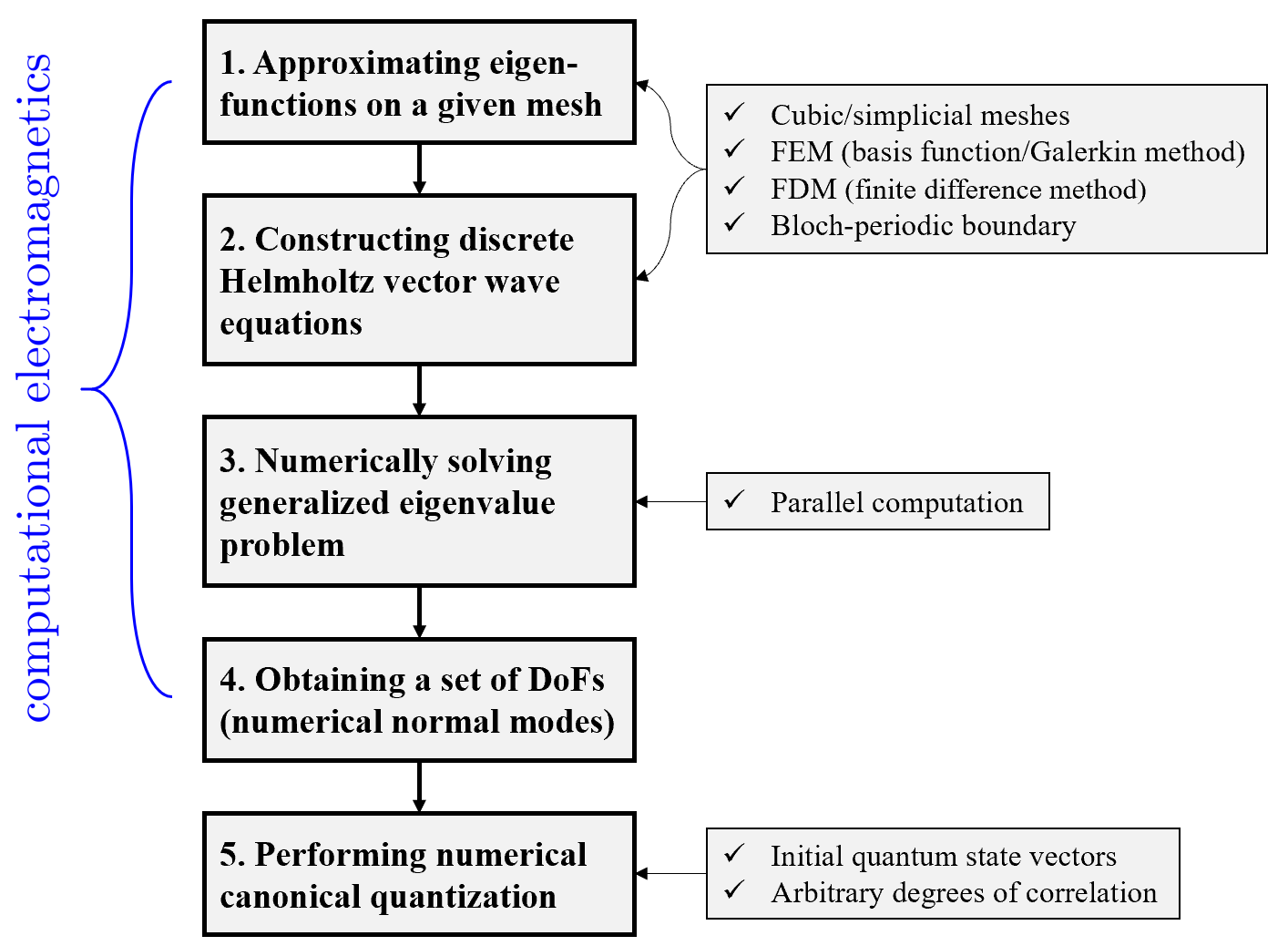}}
\caption{Schematic diagram for the procedure of numerical canonical quantization.  It is to be noted that the first four boxes are obtained using computational electromagnetics method.}
\label{fig:schm_num_quant}
\end{figure}

\section{Numerical mode decomposition and quantization}

In this section, we present the detailed procedure on how to perform numerical canonical quantization. A schematic diagram for the overall procedure of numerical canonical quantization is illustrated in Fig. \ref{fig:schm_num_quant}.

In particular, here, we are focusing on the development of a customized 1-D formulation for two-photon interference occurring in QBS to observe the Hong-Ou-Mandel (HOM) effect \cite{Hong1987measurement}, as illustrated in Fig. \ref{fig:QBS_cases}.
Note that a 2-D QBS problem with oblique incidence is mathematically homomorphic (analogous) to the 1-D problem; therefore, the HOM effect can be still observed in 1-D QBS simulations.

In what follows, curly braket $\left\{\cdot\right\}$ and square braket $\left[\cdot\right]$ denotes column vector and matrix, respectively.

\subsection{Corresponding 1-D variables}
We consider a 1-D single-layered dielectric slab in free space whose permittivity is given by
\begin{flalign}
\epsilon\left(x\right)=
\left\{
\begin{matrix}
\epsilon_{s}\epsilon_{0}, &\text{for~} \left|x\right|\leq \frac{R_{s}}{2}
\\
\epsilon_{0}, &\text{elsewhere}
\end{matrix}
\right.
.
\label{eq:original_system}
\end{flalign}
By assuming that the system of interest is in a primitive unit cell where $x\in\left[-R_{x}/2,R_{x}/2\right]$, the permittivity becomes a periodic function as
\begin{flalign}
\epsilon\left(x\right)=\epsilon\left(x+R_{x}\right).
\end{flalign}
The fact should be elaborated that although the above assumption makes the resulting physics equivalent to that of photonic crystals, capturing the local physics in the primitive unit cell within a proper time window is of our main interest.
In other words, we first look for the complete set of eigenfunctions from periodic systems and exploit them to model the original system (\ref{eq:original_system}).
It is important to note that the above assumption that the medium is periodic is a mathematical trick to numerically deal with quantum optical phenomena in the presence of inhomogeneous dielectric media but nothing more.
Note that $R_{x}$ should be chosen in such a way that the local physics of our interest should be isolated against neighbor primitive cells' aliasing effects within a properly set time window.
Specifically, $R_{x}$ can be several times larger than the localized size of photons or photon interference length.

In the 1-D periodic dielectric medium, we look for 1-D solutions of vector potential with single perpendicular polarization, i.e.,
\begin{flalign}
\tilde{\mathbf{A}}_{\boldsymbol{\kappa}}\left(\mathbf{r}\right)
=
\hat{z}\tilde{A}_{\kappa}\left(x\right)
=
\hat{z}\phi^{(\kappa)}\left(x\right)
\label{eq:1D_sol}
\end{flalign}
where the Bloch wavevector is $\boldsymbol{\kappa}=\hat{x}\kappa$.
Then, the solutions are governed by the scalar Helmholtz wave equation
\begin{flalign}
\frac{d^{2}}{dx^{2}}\phi^{(\kappa)}\left(x\right)
+
\epsilon\left(x\right)\mu_{0}\omega_{\kappa}^{2}\phi^{(\kappa)}\left(x\right)
=0.
\label{eq:1-D_helmholtz_wave_eq}
\end{flalign}
Note that the solution set of (\ref{eq:1-D_helmholtz_wave_eq}) is uncountably-infinite.
In following sections, we show how the solution set is reduced to being countably-finite with the use of a proper boundary condition and numerical methods such as finite-difference (FDM) \cite{taflove2000computational} or finite-element methods (FEM) \cite{strang2008analysis} so that one can implement a code to perform numerical experiments on quantum information propagation.

\subsection{Solving the generalized eigenvalue problem}
For a given mesh consisting of $N^{(0)}$ number of grid points, utilizing numerical methods such as FDM or FEM to (\ref{eq:1-D_helmholtz_wave_eq}), one can obtain a linear system taking form of
\begin{flalign}
\left[\mathcal{S}\right]\cdot\left\{\phi^{(\kappa)}\right\}
+
\omega^{2}_{\kappa}\mu_{0}
\left[\mathcal{M}\right]\cdot\left\{\phi^{(\kappa)}\right\}=0
\label{eq:disc_helmtholz_mat}
\end{flalign}
where $N^{(1)}=N^{(0)}-1$, $\left\{\phi^{(\kappa)}\right\}$ is an eigenvector having an eigenfrequency $\omega_{\kappa}$, sized by $N^{(1)}
\times 1$, and $\left[\mathcal{S}\right]$ and $\left[\mathcal{M}\right]$
are so-called stiffness and mass matrices in the FEM parlance, see also Appendix section \ref{app:fem} \cite{strang2008analysis}, which are Hermitian, sized by $N^{(1)} \times N^{(1)}$.
Detailed procedures to obtain (\ref{eq:disc_helmtholz_mat}) are explained in \ref{app:numerical_methods}.

Equation (\ref{eq:disc_helmtholz_mat}) corresponds to a finite-dimensional generalized eigenvalue
problem resulting from the use of numerical methods. Any standard eigensolver can be used to solve (\ref{eq:disc_helmtholz_mat}); here, we solved it using the \texttt{eig} function built in
\texttt{MATLAB}.
The eigensolver returns the total $N^{(1)}$ number of eigenvectors and eigenvalues; hence, one can eventually obtain the countably-finite solution set.
The matrix representation is
\begin{flalign}
\left[\mathcal{S}\right]\cdot\left[\Phi\right]+\left[\mathcal{M}\right]\cdot\left[\Phi\right]\cdot\left[\lambda\right]=0
\end{flalign}
where
\begin{flalign}
\left[\Phi\right]_{i,p} &=
\left\{\phi^{(\kappa_{p})}\right\}_{i},
\\
\left[\lambda\right]_{p,p} &=
\omega_{\kappa_{p}}^{2}\mu_{0},
\end{flalign}
for $p=1,2,...,N^{(1)}$.
Note that in the above representation, we label $p$-th eigenfrequency and eigenvector as $\omega_{\kappa_{p}}$ and $\left\{\phi^{(\kappa_{p})}\right\}$, respectively.
As a result, the analytic normal modes can be evaluated at grid points as
\begin{flalign}
\tilde{A}\left(x_{i},t\right)
\approx
\hat{z}\sum_{p=1}^{N_{1}}
q_{\kappa_{p}}(t)
\left[\Phi\right]_{i,p}
+\text{h.c.}
\end{flalign}

Then, the discrete version of the orthonormal relation is
\begin{flalign}
\mel
{\phi^{(\kappa_{p})}}
{\epsilon}
{\phi^{(\kappa_{p'})}}
&\equiv
\left\{\phi^{(\kappa_{p})}\right\}^{\dag}
\cdot
\left[\mathcal{M}\right]
\cdot
\left\{\phi^{(\kappa_{p'})}\right\}=\delta_{p,p'}
\end{flalign}
or
\begin{flalign}
\left[\Phi\right]^{\dag}
\cdot
\left[\mathcal{M}\right]
\cdot
\left[\Phi\right]
=\left[I\right]
\label{eq:ortho_rel_1D}
\end{flalign}
where $\left[I\right]$ denotes an identity matrix.

\subsection{Numerical canonical quantization}
Once numerical normal modes are obtained, the subsequent procedure of numerical canonical quantization is straightforward using the linear algebra.
Since the vector potential operator can be written as
\begin{flalign}
\hat{z}\hat{A}\left(x_{i},t\right)
=
\hat{z}\hat{A}^{(+)}\left(x_{i},t\right)
+
\hat{z}\hat{A}^{(-)}\left(x_{i},t\right)
\end{flalign}
where
\begin{flalign}
\hat{A}^{(+)}\left(x_{i},t\right)
&=
\sum_{p=1}^{N^{(1)}}
\sqrt{\frac{\hbar}{2\omega_{\kappa_{p}}}}
\left[\Phi\right]_{i,p}
e^{-i\omega_{{\kappa_{p}}}t}\hat{a}_{{\kappa_{p}}},
\\
\hat{A}^{(-)}\left(x_{i},t\right)
&=
\sum_{p=1}^{N^{(1)}}
\sqrt{\frac{\hbar}{2\omega_{\kappa_{p}}}}
\left[\Phi\right]^{\dag}_{p,i}
e^{i\omega_{{\kappa_{p}}}t}\hat{a}^{\dag}_{{\kappa_{p}}},
\end{flalign}
its equivalent matrix representations of quantum Hamiltonian and vector potential operators are
\begin{flalign}
\hat{H}&=
\left\{\hat{a}\right\}^{\dag}
\cdot
\left[\mathcal{D}_{H}\right]
\cdot
\left\{\hat{a}\right\},
\\
\left\{\hat{{A}}^{(+)}\left(t\right)\right\}
&=
\left[\Phi\right]
\cdot
\left[\mathcal{D}_{A}\left(t\right)\right]
\cdot
\left\{\hat{a}\right\},
\\
\left\{\hat{{A}}^{(-)}\left(t\right)\right\}
&=
\left\{\hat{a}\right\}^{\dag}
\cdot
\left[\mathcal{D}_{A}\left(t\right)\right]^{\dag}
\cdot
\left[\Phi\right]^{\dag},
\end{flalign}
where
\begin{flalign}
\left\{\hat{{A}}^{(+)}\left(t\right)\right\}_{i}
&=
\hat{{A}}^{(+)}\left(x_{i},t\right)
=\left\{\alpha_{i}\right\}\cdot\left\{\hat{a}\right\},
\\
\left\{\alpha_{i}\right\}_{p}&=\sqrt{\frac{\hbar}{2\omega_{{\kappa}_{p}}}}e^{-i\omega_{{\kappa}_{p}}t}\left[\Phi\right]_{i,p}
\\
\left\{\hat{a}\right\}_{p}&=\hat{a}_{{\kappa}_{p}},
\\
\left[\mathcal{D}_{H}\right]_{p,p}&=\hbar\omega_{{\kappa}_{p}},
\\
\left[\mathcal{D}_{A}\right]_{p,p}&=\sqrt{\frac{\hbar}{2\omega_{{\kappa}_{p}}}}e^{-i\omega_{{\kappa}_{p}}t}.
\end{flalign}

\section{Numerical example: 1-D quantum beam splitter}\label{sec:numerical_example}

\subsection{Design of 50/50 quantum beam splitter with $\pi$ phase offset}
\begin{figure}
\centering
{\includegraphics[width=\linewidth]{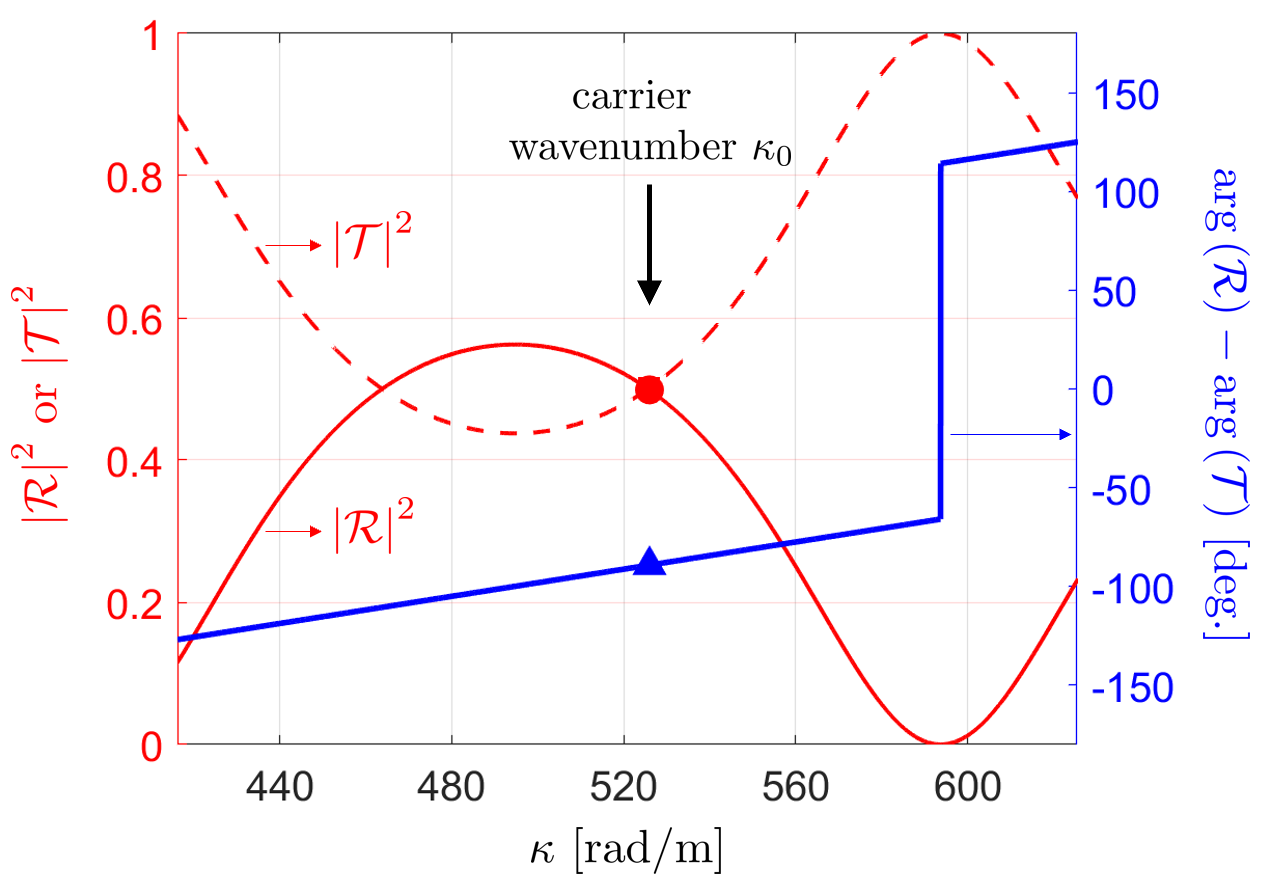}}
\caption{Reflectivity, transmittivity, and phase difference of the
designed QBS versus wavenumber.} \label{fig:QBS_design}
\end{figure}

To investigate the HOM effect, one has to design a 50/50 QBS yielding $\pi$ phase
offset between reflected and transmitted waves.
Here, performing a parametric study on reflection and transmission at a dielectric slab
\cite{BALANIS2012advanced}, we eventually set the slab permittivity
$\epsilon_{s}=7\epsilon_{0}$ [F/m] and the thickness $R_{s}=6$ [mm].
Fig. \ref{fig:QBS_design} illustrates $\left|\mathcal{R}\right|^{2}$
and $\left|\mathcal{T}\right|^{2}$ (measured on the vertical left
axis) and $\text{arg}\left(\mathcal{R}\right)-\text{arg}\left(\mathcal{T}\right)$ (measured on the
vertical right axis) over operating wavenumber.  It is to be noted that even though material dispersion is ignored, geometrical dispersion due to the finite thickness of the beam splitter is present.

 At $\kappa=\kappa_{0}=560$
[rad/m], $\left|\mathcal{R}\right|^{2}\approx 0.4987$,
$\left|\mathcal{T}\right|^{2}\approx 0.5013$,
$\text{arg}\left(\mathcal{R}\right)-\text{arg}\left(\mathcal{T}\right)\approx -89.16$ [deg.];
hence, $\kappa_{0}$ will be chosen to be a carrier wavenumber of Gaussian
or Lorentzian wave packets on which photons ride in the later use.
All parameters used in 1-D QBS simulations are shown in Table \ref{tab:parameter}.
\begin{table}
\caption{Parameter setup for 1-D QBS simulation.}
\centering
\begin{tabular}{clclcl}
\toprule
\multicolumn{2}{c}{Problem geometry} &
\multicolumn{2}{c}{Default mesh} &
\multicolumn{2}{c}{Gaussian wave packet}
\\
\cmidrule(lr){1-2}\cmidrule(lr){3-4}\cmidrule(lr){5-6}
$R_x$ & $1.5$ [m] &
$N^{(0)}$ & $2,501$ &
$x_{0}$ & $0.375$ [m]
\\
$R_{s}$ & $6$ [mm] &
$\Delta x$ & $0.6$ [mm] &
$\Delta x_{0}$ & $0.03$ [m]
\\
$\epsilon_s$ & $7\epsilon_{0}$ [F/m] &
$N^{(0)}_{s}$ & $11$ &
$\kappa_{0}$ & $526$ [rad/m]
\\
\bottomrule
\end{tabular}
\label{tab:parameter}
\end{table}
Note that $\kappa_{0}$, $x_{0}$, and $\Delta x_{0}$ are carrier wavenumber, localization position, and variance of a wave packet, respectively.

\subsection{Initializing quantum state vectors to model isolated photons}
\begin{figure*}
\centering
{\includegraphics[width=\linewidth]{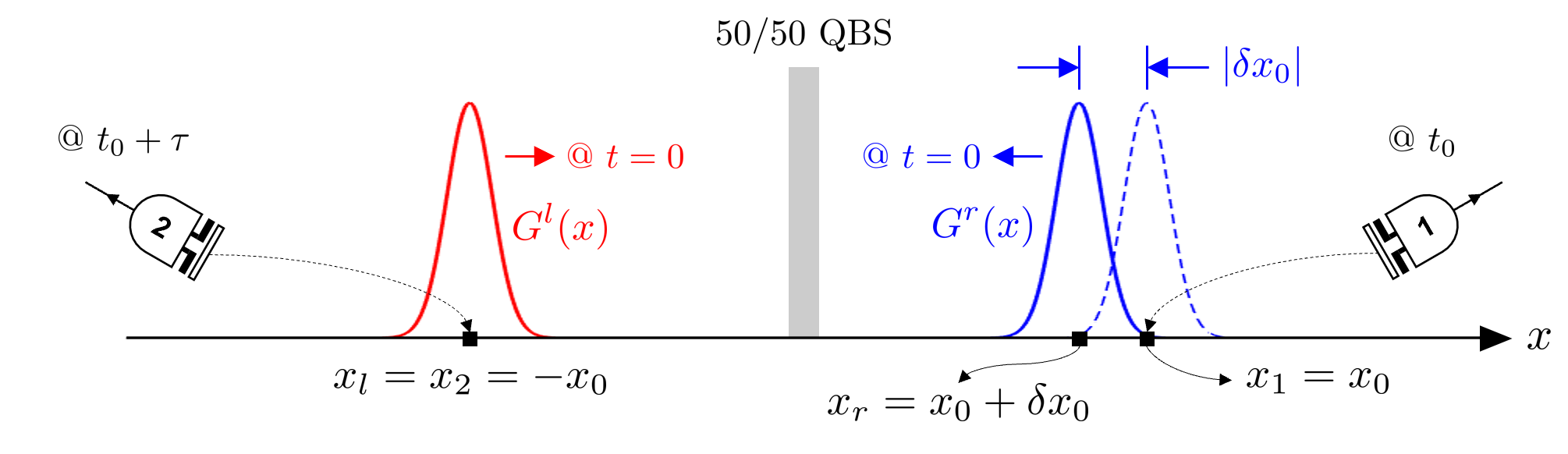}}
\caption{Schematic of the numerical experiment for two-photon incidence to observe Hong-Ou-Mandel effect.}
\label{fig:Hong_Ou_Mandel_schme}
\end{figure*}
Assume that two non-entangled photons, which are localized in space
and time in the forms of Gaussian wave packets, enter the QBS from
the left and right sides, labeled as $l$ and $r$, respectively, as
illustrated in Fig. \ref{fig:Hong_Ou_Mandel_schme}.

An initial quantum state of a single photon, localized around a specific
position in the 1-D primitive lattice, can be represented by the linear superposition of multimode Fock states.
The Fock state of one photon in $\kappa$-th normal mode can be written by $\ket{1}_{\kappa}\otimes\ket{\left\{0\right\}}$ where $\ket{\left\{0\right\}}$ is the vacuum state and its weighting factor becomes the probability amplitude, denoted by $\tilde{G}_{\kappa}$; hence, the initial quantum state vector takes the form of
\begin{flalign}
\ket{\Psi^{(1)}}
&=
\sum_{\kappa}
\tilde{G}_{\kappa}
\ket{1}_{\kappa}\otimes\ket{\left\{0\right\}}
=
\sum_{\kappa}
\tilde{G}_{\kappa}
\hat{a}_{\kappa}^{\dag}\ket{\left\{0\right\}}
\nonumber\\
&=
\left\{\hat{a}\right\}^{\dag}
\cdot
\left\{\tilde{G}\right\}
\ket{\left\{0\right\}}.
\label{eq:one-photon-quantum-state}
\end{flalign}
The above implies that a single photon is ``riding'' on top of a wave packet formed by a linear superposition of many  modes.
Note that the last equality in (\ref{eq:one-photon-quantum-state}) is the representation based on numerical canonical quantization.
The probability amplitude $\tilde{G}_{\kappa}$ is subject to the normalization condition that
\begin{align}
\sum_{\kappa}
|\tilde{G}_{\kappa}|^2=1.
\end{align}
Then, the probability of finding the photon in the $\kappa$-th mode is proportional to $|\tilde{G}_{\kappa}|^2$.
Hence, the photon does not have a definite momentum or energy but becomes momentum- or energy-uncertain  \cite{Mandel1995optical}[Chap. 12].
Since Fock states are not a function of positions, thus, the spatial
information of the single photon is incorporated into a set of
probability amplitudes.
The probability amplitude can be viewed as
{\it modal spectral amplitude}; hence, an initial spatial shape
function ($t=0$) for the single photon wave packet can be expanded by numerical
normal modes \cite{Mandel1995optical}[Chap. 12] as
\begin{flalign}
G\left(x;\kappa_{0};x_{0};\Delta x_{0}\right)
&=G_{0}
e^{-\left(\frac{x-x_{0}}{\sqrt{2}\Delta x_{0}}\right)^{2}}
e^{i \kappa_{0}\left(x-x_{0}\right)}
\nonumber \\
&=
\sum_{\kappa}
\tilde{G}_{\kappa}
\tilde{A}_{\kappa}\left(x\right)
\label{eq:FS_prob_amp}
\end{flalign}
where $G_{0}$ is a constant for $\ket{\Psi^{(1)}}$ to be normalized.

The probability amplitude $\tilde{G}_{\kappa}$ can be evaluated by the orthonormal relation (\ref{eq:ortho_rel_1D}) as
\begin{flalign}
\left\{\tilde{G}\right\}
=
\left[\Phi\right]^{\dag}
\cdot
\left[\mathcal{M}\right]
\cdot
\left\{G\right\}.
\label{eq:disc_projection}
\end{flalign}

Then, an initial quantum state for two non-entangled photons, which are localized around $x_{l}$ and $x_{r}$ (see Fig. \ref{fig:Hong_Ou_Mandel_schme}), are given by the tensor product of two individual single-photon quantum states as
\begin{flalign}
\ket{\Psi^{(2)}}
&=
\ket{\Psi^{(1)}_{l}}\otimes \ket{\Psi^{(1)}_{r}}
\nonumber \\
&=
\left\{\hat{a}\right\}^{\dag}
\cdot
\left\{\tilde{G}^{(l)}\right\}
\left\{\hat{a}\right\}^{\dag}
\cdot
\left\{\tilde{G}^{(r)}\right\}
\ket{\left\{0\right\}}
\label{eq:two_photon_quantum_state}
\end{flalign}
where, for $\nu=l$ or $r$, $\left\{\tilde{G}^{(\nu)}\right\}
=
\left[\Phi\right]^{\dag}
\cdot
\left[\mathcal{M}\right]
\cdot
\left\{G^{(\nu)}\right\}$ and $\left\{G^{(\nu)}\right\}_{i}=G\left(x_{i};\kappa_{\nu};x_{\nu};\Delta x_{\nu}\right)$.
In our simulation, it is assumed that $x_{{r}}=-x_{{l}}=x_{0}$, $\Delta x_{{r}}=\Delta x_{{l}}=\Delta x_{0}$, and $\kappa_{{l}}=-\kappa_{{r}}=\kappa_{0}$.
Then, we perturb $x_{{r}}$ by $\delta x_{0}$, viz., $x_{{r}}=x_{0}+\delta
x_{0}$ to observe the Hong-Ou-Mandel dip.

One can also assume that each photon rides on a Lorentzian
wave packet whose spatial distribution takes the form of double-sided
exponential or Laplace distribution given by
\begin{flalign}
L\left(x;\kappa_{\nu};x_{\nu};\Delta x_{\nu}\right)
&=
L_{0}
e^{-\frac{\left|x-x_{\nu}\right|}{\Delta x_{\nu}}}
e^{i \kappa_{\nu}\left(x-x_{\nu}\right)}.
\end{flalign}
Then, the corresponding initial quantum state can be obtained similar to (\ref{eq:two_photon_quantum_state}).

\subsection{Second order correlation function and HOM dip}
Indistinguishability of two photons can be measured through the
degree of intensity coherence, called
second order correlation, denoted as $g^{(2)}$.
It was first reported in \cite{Hanbury1957interferometry} for the development of advanced stellar interferometers with classical light beam inputs.
The quantum version of the second order correlation function can be deduced from the physical mechanism of the simultaneous detection of two photons  \cite{Glauber1963quantum}.
In other words, the action of an annihilation operator on a quantum
state resembles absorption of a single photon at a specific location
and certain time instant, viz., the photoelectric effect.
The detailed derivation for the quantum second order correlation function can be found in \cite{Mandel1995optical}[Chap. 12].  Also,
$g^{(2)}$ is widely used in many quantum-associated experiments to test the indistinguishability of two photons.

We place two photodetectors at $x_{1}$ and $x_{2}$ and each
photodetector is supposed to detect a photon at time instant $t_{0}$ and
$t_{0}+\tau$, respectively.
Note that $t_{0}=2x_{0}/c$ and $\tau=\delta x_{0}/c$.
Therefore, corresponding $g^{(2)}$ is a function of $\tau$ only and can be written as
\begin{flalign}
g^{(2)}\left(x_{1},t_{0};x_{2},t_{0}+\tau\right)
=
\frac{A}{B_{1}B_{2}}
\label{eq:g_2_func}
\end{flalign}
where $\tau$ is a free variable,
\begin{flalign}
A&=
{\bra{\Psi^{(2)}}}
{
\hat{A}^{(-)}\left(x_{1},t_{0}\right)
\hat{A}^{(-)}\left(x_{2},t_{0}+\tau\right)
}
\nonumber \\
&\times
{
\hat{A}^{(+)}\left(x_{2},t_{0}+\tau\right)
\hat{A}^{(+)}\left(x_{1},t_{0}\right)
}
{\ket{\Psi^{(2)}}}
,
\\
&\!\!\!\!\!
B_{1}=
{\bra{\Psi^{(2)}}}
{
\hat{A}^{(-)}\left(x_{2},t_{0}\right)
\hat{A}^{(+)}\left(x_{2},t_{0}\right)
}
{\ket{\Psi^{(2)}}}
,
\\
&\!\!\!\!\!
B_{2}=
{\bra{\Psi^{(2)}}}
{
\hat{A}^{(-)}\left(x_{2},t_{0}+\tau\right)
\hat{A}^{(+)}\left(x_{2},t_{0}+\tau\right)
}
{\ket{\Psi^{(2)}}}
,
\label{eq:g_two_func_two_photon}
\end{flalign}
and by the time order
\begin{flalign}
x_{1}&=\left\{
\begin{matrix}
x_{r},&\text{for}~\tau\geq0\\
x_{l},&\text{for}~\tau\leq0\\
\end{matrix}
\right.
,
\\
x_{2}&=\left\{
\begin{matrix}
x_{l},&\text{for}~\tau\geq0\\
x_{r},&\text{for}~\tau\leq0\\
\end{matrix}
\right.
.
\end{flalign}
The detailed procedure to evaluate the above second order correlation function is explained in \ref{ap:sec_cor}.

\begin{figure}
\centering
{\includegraphics[width=\linewidth]{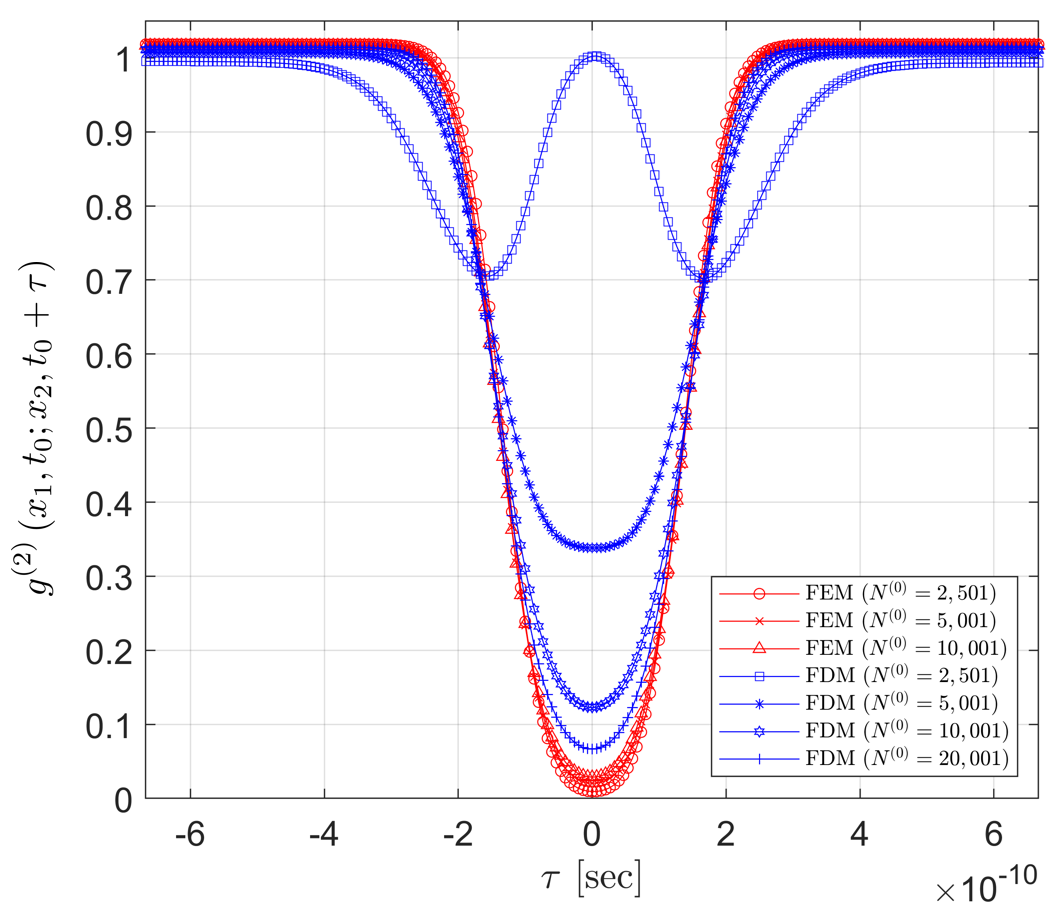}}
\caption{The Hong-Ou-Mandel effect is numerically evaluated.
Two sets of curves colored in red and blue illustrate results by FDM-
and FEM-solvers, respectively, for different meshes.}
\label{fig:Hong_Ou_Mandel_Curve}
\end{figure}

Fig. \ref{fig:Hong_Ou_Mandel_Curve} depicts
$g^{(2)}\left(x_{1},t_{0};x_{2},t_{0}+\tau\right)$ versus $\tau$.
As expected, when $\tau=0$, there is the HOM dip approaching to almost zero since the pair of photons are perfectly identical; hence, they are indistinguishable. From FDM and FEM results with the most refined meshes, the average visibility
of the HOM dip was recorded as $95.24\%\pm1.89\%$.
Furthermore, overall shapes of our numerically-evaluated HOM curves look Gaussian, which agree with the theoretical explanation \cite{Rohde2005optimal} that the power spectrum of the single-photon wave packet is strongly relevant to patterns of HOM curves.

Note that despite
the perfect identicality of the two photons, the
non-perfect-visibility is due to the narrow bandwidth
characteristics of the single dielectric slab.
It is seen that FEM
results (red curves) have the faster convergence than FDM results
(blue curves) since the former method accurately captures the abrupt change
of the medium inhomogeneity around slab interfaces.

It should be
mentioned that the HOM dip, shown in Fig.
\ref{fig:Hong_Ou_Mandel_Curve}, is usually less than one half, as
evident in experimental results. This is entirely a quantum effect
unachievable by solving classical Maxwell's equation
that produces the dip always greater than one half depending on the
type of incident fields such as coherent pulses or chaotic lights
\cite{Facao2011computer,Kim2013conditions}.

\begin{figure}
\centering
{\includegraphics[width=\linewidth]{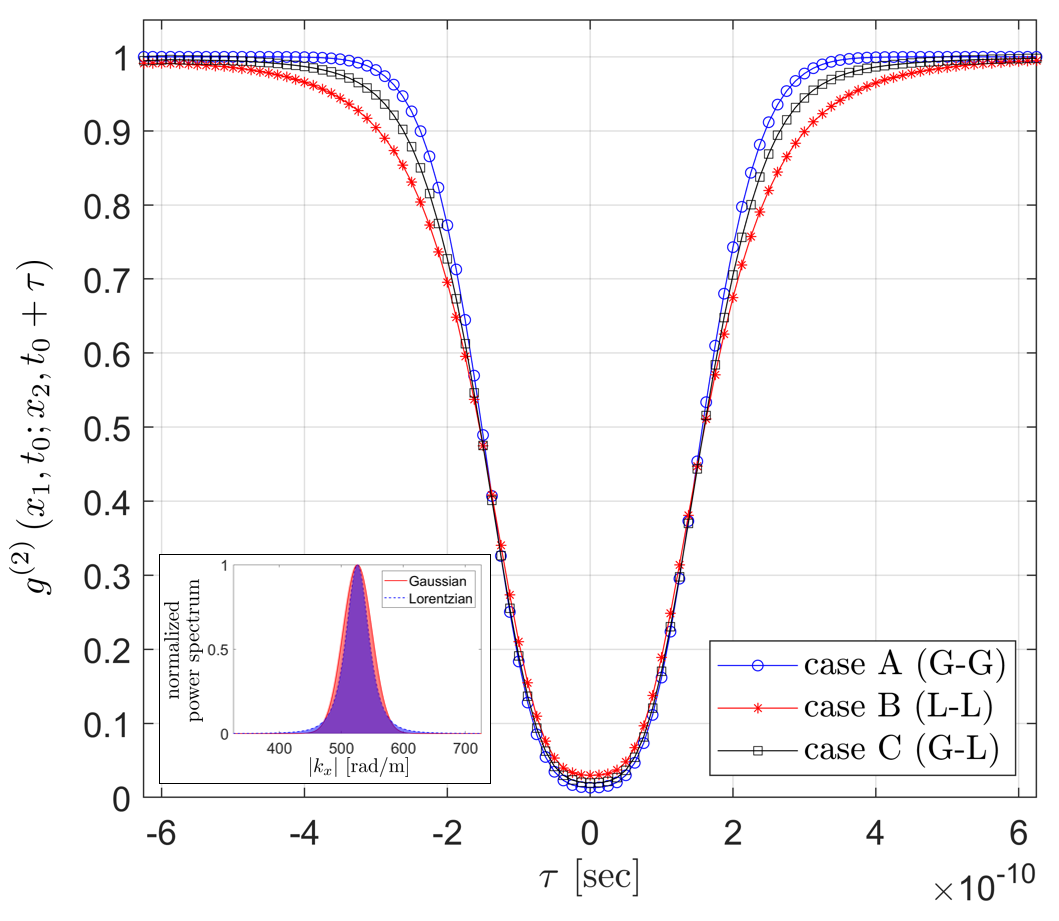}}
\caption{$g^{(2)}_{ta}\left(\tau\right)$ versus $\tau$ for different
wave packets of two incoming photons: Gaussian and Gaussian (case A),
Lorentzian and Lorentzian (case B), and Gaussian and Lorentzian
(case C).} \label{fig:Hong_Ou_Mandel_Curve_2}
\end{figure}
To observe effects of the HOM dip when wave packets of two incoming
photons are non-identical but highly-correlated with respect to
their power spectrum, we study $g^{(2)}\left(x_{1},t_{0};x_{2},t_{0}+\tau\right)$ when left and right photons ride on Gaussian and Lorentzian wave packets,
respectively, depicted in Fig. \ref{fig:Hong_Ou_Mandel_Curve_2}. Two
photons are riding on Gaussian and Gaussian wave packets in case A,
Lorentzian and Lorentzian wave packets in case B, and Gaussian and
Lorentzian wave packets in case C, respectively. All results are
obtained by using FEM-solvers with the default mesh
($N^{(0)}=2,501$). Here, it is assumed that both wave packets have
the same variance $\Delta x_{0}$ and carrier wavenumber. Their
normalized power spectrum is displayed in the inset of Fig.
\ref{fig:Hong_Ou_Mandel_Curve_2}. It can be seen that the HOM dip
still exists in case C and its visibility is comparable to the
others (identical wave packets). Again, the indistinguishability of a
pair of photons originates from the high correlation in power
spectrum of their wave packets but not identicality.

\section{Conclusion}
We have used numerical canonical quantization to solve quantum Maxwell's equations based on: (1) the macroscopic (phenomenological) electromagnetic theory on quantum electrodynamics, and (2) concepts borrowed from computational electromagnetics.
Instead of {\it ad-hoc} analytic normal modes, we numerically constructed complete set of and time-reversible normal modes in the form of traveling waves to diagonalize the Hamiltonian.
Namely, we directly solved the Helmholtz wave equations for a general linear, reciprocal, isotropic, non-dispersive, and inhomogeneous dielectric media by using either finite-element or finite-difference methods.
With the use of {\it Bloch-periodic boundary conditions}, and numerical procedure, we eventually converted a scattering problem with uncountably-infinite number of modes into one with a countably-finite number of modes.
Those modes corresponded to numerical Bloch-Floquet-like normal modes sparsely-sampled at a specific Bloch wavenumber.
Therefore, the procedure of numerical canonical quantization was straightforward using linear algebra.
We provided relevant numerical recipes in detail and showed the important numerical example of indistinguishable two-photon interference in quantum beam splitters, exhibiting Hong-Ou-Mandel effect, which is purely a quantum effect.

The {\it numerical canonical quantization} involves two processes: (1) to diagonalize the classical Maxwell's equations or its ensuing Hamiltonian using classical CEM (computational electromagnetics) methods; (2) The quantization of the modes by the elevation of the field or conjugate variables to become quantum operators.  The subsequent quantum operators are diagonalized by the photon number or Fock's state, which  are well known.  The double diagonalizations allow the ease of propagating quantum information on a numerical grid.  It is extendable to multidimensional cases in the future, which will serve as numerical experiments to study future quantum phenomena.
Hence, it would eventually allow quantum-optical numerical experiments of high fidelity to replace many real experiments.
As such, through numerical experiments, one can explore a large variety of quantum-optical phenomena in complex systems that still perplex the community: cases in point are virtual photons, quantum-optical metamaterials, superluminal photons by quantum tunneling, interaction free measurement, NOON states in quantum metrology and quantum sensing, and time entanglement.

Also, we introduce the concept of photons riding on a wave packet in developing our numerical procedure.  It helps to guide us correctly through our numerical work, and it is corroborated by predicting the HOM effect correctly, which has been observed in the laboratory.

This work involves entirely linear systems; although photons do not interact with each other in such a system, they interfere with each other.  Hence, it is still useful to study quantum effects of multiple photon interference in light-matter interaction.  Future involvement of nonlinear optics and nonlinear effects will more richly endow these numerical experiments.

It is to be noted that the fully atomistic nature of matter is not accounted for in this work.  The coupling of the harmonic oscillators to the vast degree of freedom in the atomistic model will give rise to quantum losses and emergence of Langevin sources.  This is discussed in \cite{Sha2018dissipative,SCHEEL2008numerical,Dung1998three,huttner1992quantization}.

\begin{acknowledgments}
The work is funded by NSF 1818910 award and a startup fund at Purdue university.
WCC is also funded by DVSS at HKU, Summer 2019.
\end{acknowledgments}

\appendix
\section{Calculation of degree of quantum coherence}\label{ap:sec_cor}
In \ref{ap:sec_cor}, we provide a specific numerical recipe on how to calculate arbitrary degrees of quantum coherence eventually taking the form of
\begin{flalign}
\mel
{0}
{\hat{O}}
{0}
\end{flalign}
where an arbitrary operator ${\hat{O}}$ consists of products of the weighted sum of annihilation or creation operators.
For example, the first order correlation for a single-photon quantum state vector, which is related to the detection probability of the photon at specific position and time instant or the expectation value of the energy density, can be written as
\begin{flalign}
\hat{O}
&=
\left(
\sum_{p'''}\gamma^{*}_{p'''}\hat{a}_{\kappa_{p'''}}
\right)
\left(
\sum_{p''}\alpha^{*}_{p''}\hat{a}^{\dag}_{\kappa_{p''}}
\right)
\nonumber\\
&\times
\left(
\sum_{p'}\alpha_{p'}\hat{a}_{\kappa_{p'}}
\right)
\left(
\sum_{p}\gamma_{p}\hat{a}_{\kappa_{p}}^{\dag}
\right)
\end{flalign}
where $\alpha$ and $\gamma$ are constants associated with the vector potential operator and quantum state vector.

The principle of the present method are based on (1) {\it reordering process} and (2) {\it constantization process}.
Note that, in \ref{ap:sec_cor}, a curly braket $\left\{\cdot\right\}$ denotes a column vector and $\cdot$ is used for the matrix multiplication.

\subsection{Reordering process}
Consider two operators $\hat{A}$ and $\hat{B}$ which are given by
\begin{flalign}
\hat{A}&=\sum_{p}\alpha_{p}\hat{a}_{\kappa_{p}}=\left\{\alpha\right\}^{t}\cdot\left\{\hat{a}\right\},
\\
\hat{B}&=\sum_{p}\beta_{p}\hat{a}^{\dag}_{\kappa_{p}}=\left\{\hat{a}\right\}^{\dag}
\cdot
\left\{\beta\right\}
\end{flalign}
where $\left\{\hat{a}\right\}$, $\left\{\alpha\right\}$, and $\left\{\beta\right\}$ denote column vectors whose elements are
\begin{flalign}
\left\{\hat{a}\right\}_{p}=\hat{a}_{\kappa_{p}},\quad \left\{\hat{a}\right\}^{\dag}_{p}&=\hat{a}^{\dag}_{\kappa_{p}},
\\
\left\{\alpha\right\}_{p}=\alpha_{p}, \quad \left\{\beta\right\}_{p}&=\beta_{p}.
\end{flalign}
Then, their product $\hat{B}\hat{A}$ is
\begin{flalign}
\hat{B}\hat{A}
&=
\left(
\sum_{p}\beta_{p}\hat{a}^{\dag}_{\kappa_{p}}
\right)
\left(
\sum_{p'}\alpha_{p'}\hat{a}_{\kappa_{p'}}
\right)
\nonumber\\
&=
\sum_{p,p'}\beta_{p}\alpha_{p'}\hat{a}^{\dag}_{\kappa_{p}}\hat{a}_{\kappa_{p'}}.
\label{eq:ba_prod}
\end{flalign}
By using the commutator relation, ($\ref{eq:ba_prod}$) can be also expressed as
\begin{flalign}
\hat{B}\hat{A}
&=\sum_{p,p'}\beta_{p}\alpha_{p'}\left(\hat{a}_{\kappa_{p'}}\hat{a}^{\dag}_{\kappa_{p}}-\delta_{p,p'}\hat{I}\right)
\nonumber\\
&=\sum_{p,p'}\alpha_{p'}\beta_{p}\hat{a}_{\kappa_{p'}}\hat{a}^{\dag}_{\kappa_{p}}
-
\sum_{p}\alpha_{p}\beta_{p}\hat{I}
\nonumber\\
&=
\left(
\sum_{p'}\alpha_{p'}\hat{a}_{\kappa_{p'}}
\right)
\left(
\sum_{p}\beta_{p}\hat{a}^{\dag}_{\kappa_{p}}
\right)
-
\sum_{p}\alpha_{p}\beta_{p}\hat{I}
\nonumber\\
&=
\hat{A}\hat{B}
-
\sum_{p}\alpha_{p}\beta_{p}\hat{I}
.
\label{eq:ba_prod_2}
\end{flalign}
One can also rewrite (\ref{eq:ba_prod_2}) in the equivalent matrix representation as
\begin{flalign}
\hat{B}\hat{A}
&=
\left\{\hat{a}\right\}^{\dag}
\cdot
\left\{\beta\right\}
\left\{\alpha\right\}^{t}
\cdot
\left\{\hat{a}\right\}
\nonumber \\
&=
\left\{\alpha\right\}^{t}
\cdot
\left\{\hat{a}\right\}
\left\{\hat{a}\right\}^{\dag}
\cdot
\left\{\beta\right\}
-
\left\{\alpha\right\}^{t}\cdot\left\{\beta\right\}\hat{I}
\nonumber \\
&=
\hat{A}
\hat{B}
-
\left\{\alpha\right\}^{t}
\cdot
\left\{\beta\right\}\hat{I}.
\label{eq:convert_ladder_op_order}
\end{flalign}

\subsection{Constantization process}
Any successive action of a set of  creation and annihilation operators to the vacuum state can be replaced by a constant.
Consider the product of operators $\hat{A}\hat{B}$ acting on the vacuum, then,
\begin{flalign}
\hat{A}\hat{B}\ket{0}
&=
\left(
\sum_{p}\alpha_{p}\hat{a}_{\kappa_{p}}
\right)
\left(
\sum_{p'}\beta_{p'}\hat{a}^{\dag}_{\kappa_{p'}}
\right)
\nonumber \\
&=
\sum_{p,p'}\alpha_{p}\beta_{p'}\hat{a}_{\kappa_{p}}
\hat{a}^{\dag}_{\kappa_{p'}}
\ket{0}
=
\sum_{p}
\alpha_{p}
\beta_{p}
\ket{0}
\nonumber\\
&=
\left\{\alpha\right\}^{t}
\cdot
\left\{\hat{a}\right\}
\left\{\hat{a}\right\}^{\dag}
\cdot
\left\{\beta\right\}
\ket{0}
=
\left\{\alpha\right\}^{t}\cdot\left\{\beta\right\}
\ket{0}
\label{eq:ann_cre_vac_rel}
\end{flalign}
since
\begin{flalign}
\hat{a}_{\kappa_{p}}\hat{a}^{\dag}_{\kappa_{p'}}\ket{0}
=
\delta_{p,p'}\ket{0}.
\end{flalign}

\subsection{First and second order correlations for two-photon quantum state vector}
For example, the first order correlation for two photons is calculated by
\begin{widetext}
\begin{flalign}
A
&=
\mel
{\Psi^{(2)}}
{
\hat{{A}}^{(-)}\left(x_{i},t\right)
\hat{{A}}^{(+)}\left(x_{i},t\right)
}
{\Psi^{(2)}}
\nonumber \\
&=
\bra{0}
\left\{\beta_{1}\right\}^{\dag}\cdot\left\{\hat{a}\right\}
\underbrace{
\left\{\beta_{2}\right\}^{\dag}\cdot\left\{\hat{a}\right\}
\left\{\hat{a}\right\}^{\dag}\cdot\left\{\alpha\right\}^{*}
}_{\text{use~(\ref{eq:convert_ladder_op_order})}}
\underbrace{
\left\{\alpha\right\}^{t}\cdot\left\{\hat{a}\right\}
\left\{\hat{a}\right\}^{\dag}\cdot\left\{\beta_{2}\right\}
}_{\text{use~(\ref{eq:convert_ladder_op_order})}}
\left\{\hat{a}\right\}^{\dag}\cdot\left\{\beta_{1}\right\}
\ket{0}
\nonumber \\
&=
\bra{0}
\left\{\beta_{1}\right\}^{\dag}\cdot\left\{\hat{a}\right\}
\left(
\left\{\hat{a}\right\}^{\dag}\cdot\left\{\alpha\right\}^{*}
\left\{\beta_{2}\right\}^{\dag}\cdot\left\{\hat{a}\right\}
-
\left\{\beta_{2}\right\}^{\dag}\cdot\left\{\alpha\right\}^{*}\hat{I}
\right)
\left(
\left\{\hat{a}\right\}^{\dag}\cdot\left\{\beta_{2}\right\}
\left\{\alpha\right\}^{t}\cdot\left\{\hat{a}\right\}
-
\left\{\alpha\right\}^{t}\cdot\left\{\beta_{2}\right\}\hat{I}
\right)
\left\{\hat{a}\right\}^{\dag}\cdot\left\{\beta_{1}\right\}
\ket{0}
\nonumber \\
&\underbrace{=}_{\text{use (\ref{eq:ann_cre_vac_rel})}}
\left\{\beta_{1}\right\}^{\dag}\cdot\left\{\alpha\right\}^{*}
\left\{\beta_{2}\right\}^{\dag}\cdot\left\{\beta_{2}\right\}
\left\{\alpha\right\}^{t}\cdot\left\{\beta_{1}\right\}
-
\left\{\alpha\right\}^{t}\cdot\left\{\beta_{1}\right\}
\left\{\beta_{1}\right\}^{\dag}\cdot\left\{\beta_{2}\right\}
\left\{\beta_{2}\right\}^{\dag}\cdot\left\{\alpha\right\}^{*}
\nonumber \\
&-
\left\{\beta_{1}\right\}^{\dag}\cdot\left\{\alpha\right\}^{*}
\left\{\beta_{2}\right\}^{\dag}\cdot\left\{\beta_{1}\right\}
\left\{\alpha\right\}^{t}\cdot\left\{\beta_{2}\right\}
+
\left\{\alpha\right\}^{t}\cdot\left\{\beta_{2}\right\}
\left\{\beta_{1}\right\}^{\dag}\cdot\left\{\beta_{1}\right\}
\left\{\beta_{2}\right\}^{\dag}\cdot\left\{\alpha\right\}^{*}.
\label{eq:fir_corr}
\end{flalign}
\end{widetext}
The above algebra simplifies because of the second stage diagonalization:  The photon number state, in this case, the vacuum state, is the eigenstate of the quantum operators.
And the second order correlation for two photons is
\begin{flalign}
A
&=
\braket{A'}
\end{flalign}
where
\begin{widetext}
\begin{flalign}
\ket{A'}
&=
{
\hat{{A}}^{(+)}\left(x_{j},t\right)
\hat{{A}}^{(+)}\left(x_{i},t\right)
}
\ket{\Psi^{(2)}}
=
\left\{\alpha_{j}\right\}^{t}\cdot\left\{\hat{a}\right\}
\underbrace{
\left\{\alpha_{i}\right\}^{t}\cdot\left\{\hat{a}\right\}
\left\{\hat{a}\right\}^{\dag}\cdot\left\{\beta_{2}\right\}
}_{\text{use~(\ref{eq:convert_ladder_op_order})}}
\left\{\hat{a}\right\}^{\dag}\cdot\left\{\beta_{1}\right\}
\ket{0}
\nonumber \\
&=
\left\{\alpha_{j}\right\}^{t}\cdot\left\{\hat{a}\right\}
\left(
\left\{\hat{a}\right\}^{\dag}\cdot\left\{\beta_{2}\right\}
\left\{\alpha_{i}\right\}^{t}\cdot\left\{\hat{a}\right\}
-
\left\{\alpha_{i}\right\}^{t}\cdot\left\{\beta_{2}\right\}\hat{I}
\right)
\left\{\hat{a}\right\}^{\dag}\cdot\left\{\beta_{1}\right\}
\ket{0}
\nonumber \\
&\underbrace{=}_{\text{use (\ref{eq:ann_cre_vac_rel})}}
\left(
\left\{\alpha_{j}\right\}^{t}\cdot\left\{\beta_{2}\right\}
\left\{\alpha_{i}\right\}^{t}\cdot\left\{\beta_{1}\right\}
-
\left\{\alpha_{i}\right\}^{t}\cdot\left\{\beta_{2}\right\}
\left\{\alpha_{j}\right\}^{t}\cdot\left\{\beta_{1}\right\}
\right)
\ket{0}.
\end{flalign}
\end{widetext}

\section{Numerical methods}\label{app:numerical_methods}
\subsection{Mesh generation}
\begin{figure*}
\centering
\includegraphics[width=\linewidth]{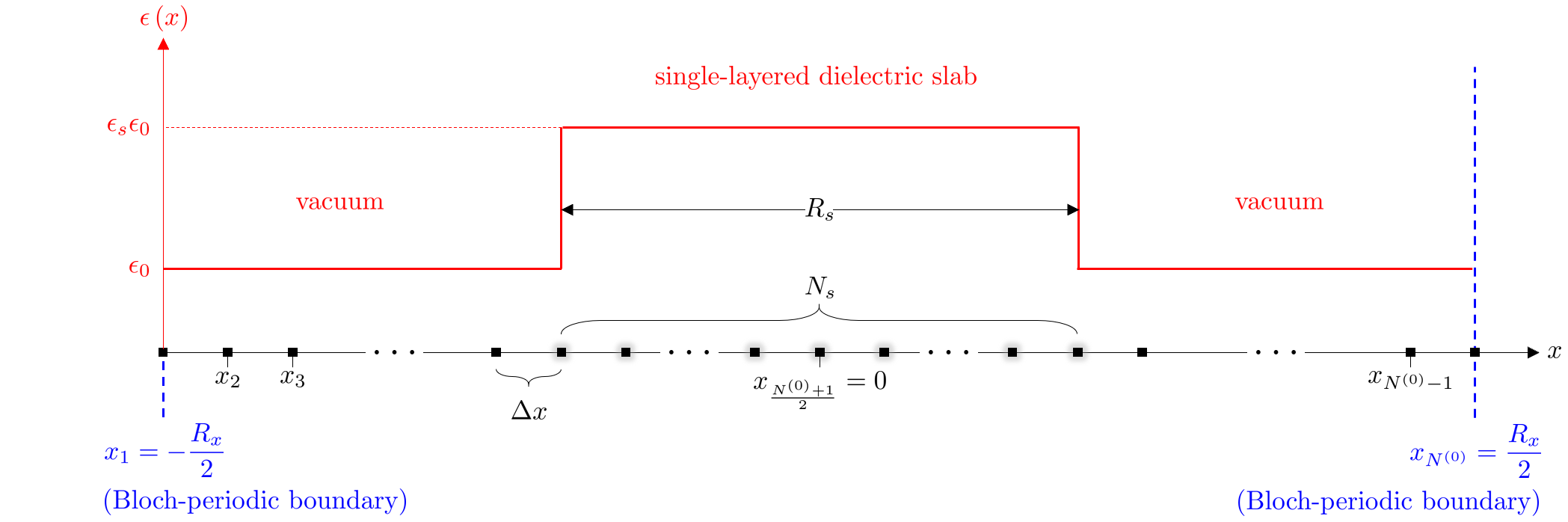}
\caption{Schematic of 1-D mesh.} \label{fig:mesh}
\end{figure*}

Consider a set of discrete grid points, called mesh, evenly-spaced by $\Delta x$,
that approximates the primitive unit cell
$\Omega=\left\{x\in\left[-0.5R_{x},0.5R_{x}\right]\right\}$ by
\begin{flalign}
\Omega\approx \bigoplus_{i=1}^{N^{(0)}}x_{i}
\end{flalign}
where $x_{i}$ denotes a coordinate of $i$-th grid
point and $N^{(0)}$ and $N^{(1)}=N^{(0)}-1$ are the total number of grid points and lines, respectively, as
illustrated in Fig. \ref{fig:mesh}.
\begin{figure}
\centering
\subfloat[Global view illustrating the physics of our interest.]
{\includegraphics[width=\linewidth]{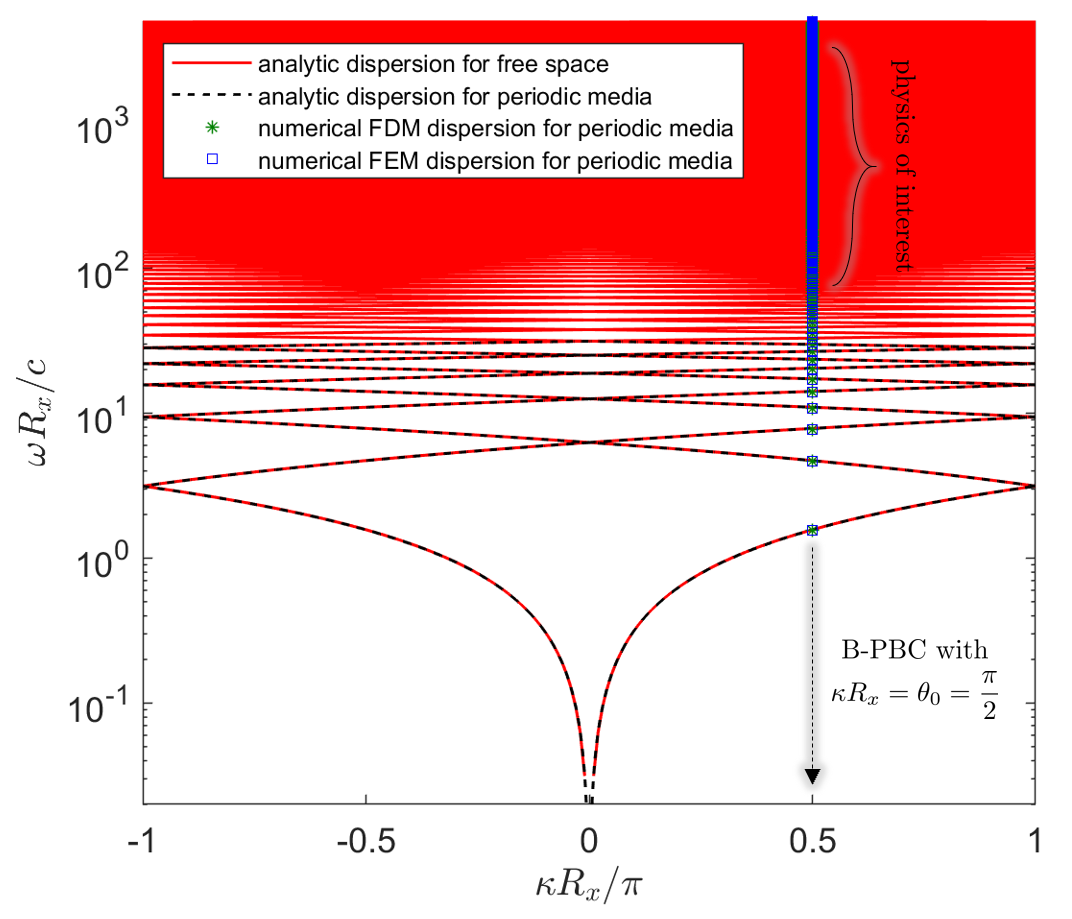}
\label{fig:dispersion}
}
\\
\subfloat[Local view at the first bandgap where the degeneracy is broken.]
{\includegraphics[width=\linewidth]{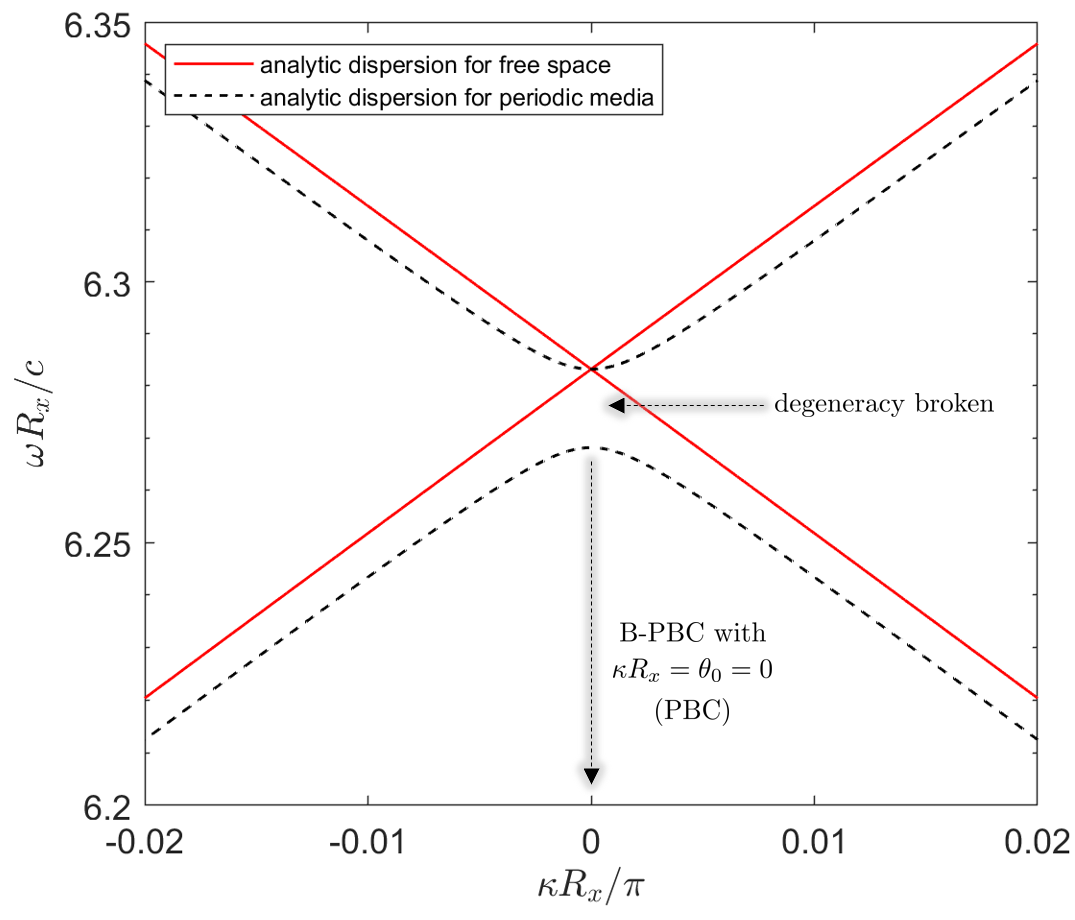}
\label{fig:degeneracy_broken}
}
\caption{Dispersion of numerical Bloch-Floquet normal modes in the first Brillouin zone.}
\label{fig:numerical_mode_dispersion}
\end{figure}

\subsection{Bloch-periodic boundary conditions}
Importantly, if one uses usual periodic boundary conditions (PBC) in the presence of inhomogeneous media, the resulting numerical normal modes become neither degenerate nor in the form of traveling waves.
It can be physically explained by the fact that wave propagation in photonic crystal is not allowed at edges of passbands.
Equivalently, the solution Bloch wavevector exactly equals to integer multiples of the reciprocal lattice vector (see Fig. \ref{fig:degeneracy_broken}).
Hence, one should apply B-PBC to obtain correct numerical normal modes used in numerical canonical quantization.
The B-PBC gets to sparsely sample analytic Bloch-Floquet normal modes at an arbitrarily displaced point $\kappa R_{x} = \theta_{0}$ in the first Brillouin zone with different eigenfrequencies (see Fig. \ref{fig:dispersion}).
This makes the solution set become countably-infinite.
The use of B-PBC consists of (1) value- and (2) derivative-periodic parts.
By the Bloch theorem, the former relates solutions at boundary grid points in such a way that
\begin{flalign}
\phi^{(\kappa)}\left(x_{N^{(0)}}=\frac{R_{x}}{2}\right)=\phi^{(\kappa)}\left(x_{1}=-\frac{R_{x}}{2}\right)e^{i \theta_0} \label{eq:1D_B-PBC}
\end{flalign}
with an arbitrary phase factor $\kappa R_{x}=\theta_{0}$ [rad].
Note that usual PBC is a special kind of B-PBC when $\theta_{0}=0$ [rad].
The latter is implicitly assumed as Neumann boundary conditions in numerical schemes.

\subsection{Finite-difference method}
Finite-difference method (FDM) \cite{taflove2000computational} is a popular numerical scheme used in a variety of scientific and
engineering areas due to its simplicity in formulation and
implementation.
Approximating second order differential operators by the central difference method and sampling solutions at every grid point, one can obtain a discrete counterpart of
(\ref{eq:1-D_helmholtz_wave_eq}) as
\begin{flalign}
\frac{\phi^{(\kappa)}_{i+1}-2\phi^{(\kappa)}_{i}+\phi^{(\kappa)}_{i-1}}{\Delta x^{2}}
+
\epsilon_{i}\mu_{0}\omega_{\kappa}^{2}\phi^{(\kappa)}_{i}=0,
\label{eq:FDM_helmtholz}
\end{flalign}
centered on $i$-th grid point for
$i=1,2,\cdots,N^{(0)}$, where $\phi^{(\kappa)}_{i}=\phi^{(\kappa)}\left(x_{i}\right)$ and $\epsilon_{i}=\epsilon\left(x_{i}\right)$ while solutions at
boundary grid points are linked in such a way that
\begin{flalign}
\phi^{(\kappa)}_{N^{(0)}}&=\phi^{(\kappa)}_{1}e^{i \theta_{0}}.
\end{flalign}
The use of B-PBC results in the decrease of degrees of freedom (DoF)
by $1$ from total number of grid points, viz., $N^{(1)}$.
To check the derivative boundary condition being periodic, one can calculate the spatial derivative of solutions using the central difference and averaging methods at the both boundaries as
\begin{flalign}
\frac
{d\phi^{(\kappa)}\left(x\right)}
{dx}
\Biggr|_{x={x_{1}}}
&\approx
\frac
{\phi^{(\kappa)}_{2}-\phi^{(\kappa)}_{0}}
{2\Delta x},
\\
\frac
{d\phi^{(\kappa)}\left(x\right)}
{dx}
\Biggr|_{x={x_{N_{0}}}}
&\approx
\frac
{\phi^{(\kappa)}_{N_{0}+1}-\phi^{(\kappa)}_{N_{0}-1}}
{2\Delta x}
=
e^{i \theta_{0}}
\frac
{\phi^{(\kappa)}_{2}-\phi^{(\kappa)}_{0}}
{2\Delta x}.
\end{flalign}
Hence,
\begin{flalign}
\frac
{d\phi^{(\kappa)}\left(x\right)}
{dx}
\Biggr|_{x={x_{N_{0}}}}
=
e^{i \theta_{0}}
\frac
{d\phi^{(\kappa)}\left(x\right)}
{dx}
\Biggr|_{x={x_{1}}}.
\end{flalign}
Finally, equivalent matrix
representation of (\ref{eq:FDM_helmtholz}) can be written as
\begin{flalign}
\left[\mathcal{S}\right]\cdot\left\{\phi^{(\kappa)}\right\}
+
\omega^{2}_{\kappa}\mu_{0}
\left[\mathcal{M}\right]\cdot\left\{\phi^{(\kappa)}\right\}=0
\label{eq:FDM_helmtholz_mat}
\end{flalign}
$\left\{\phi^{(\kappa)}\right\}$ is an eigenvector for an eigenfrequency $\omega_{\kappa}$, sized by $N^{(1)}
\times 1$, and $\left[\mathcal{S}\right]$ and $\left[\mathcal{M}\right]$
are Hermitian matrices, sized by $N^{(1)} \times N^{(1)}$, whose
elements are given by
\begin{flalign}
&\left\{\phi^{(\kappa)}\right\}_{i}= \phi^{(\kappa)}_{i},
\\
&\left[\mathcal{S}\right]_{i,j}
=
\left\{
\begin{matrix}
-2/\Delta x,
& \text{for~}j=i
\\
e^{\left(-i\right)^{n^{\pm}}\theta_{0}}/\Delta x,
& \text{for~}j=m^{\pm}
\end{matrix}
\right.
\\
&\left[\mathcal{M}\right]_{i,i} =
\epsilon_{i}\Delta x,
\end{flalign}
for $i=1,2,\cdots,N^{(1)}$. Note that, at given $i$, $m^{\pm}=i \pm
1+n^{\pm}N^{(1)}$ should be an element in the integer set of
$\left\{1,2,\cdots,N^{(1)}\right\}$ by properly choosing $n^{\pm}$.
For example, if $i=1$,
$\left[\mathcal{S}\right]_{1,N^{(1)}}=e^{-i\theta_{0}}$
where $m^{-}=N^{(1)}$ and $n^{-}=1$.

\subsection{Finite-element method}\label{app:fem}
Finite-element method (FEM) \cite{strang2008analysis} is a subspace projection methods extensively used in wave
scattering problems involving complex geometries. It approximates a
continuous solution by the weighted sum of a finite set of basis
functions as
\begin{flalign}
\phi^{(\kappa)}\left(x\right)\approx\sum_{i=1}^{N^{(0)}}\phi^{(\kappa)}_{i}W^{(0)}_{i}\left(x\right)
\label{eq:field_FEM_rep}
\end{flalign}
where $\phi^{(\kappa)}_{i}$ and $W^{(0)}_{i}\left(x\right)$ are unknowns and a
(scalar) basis function for $i$-th grid point,
respectively \footnote{This basis function is often called {\it
Whitney 0-forms} or {\it nodal elements} \cite{bossavit1988whitney}
used to expand 0-forms. In fact, in the language of differential
forms \cite{Warnick1997teaching,Teixeira1999lattice}, vector
potential $\mathcal{A}$ is 1-form but in this example its component
$A_{z}$, which is a scalar function, is modeled as 0-form, hence,
resulting $\mathcal{A}=dz\wedge A_{z}$ still remains 1-form while
$A_{z}$ being extruded by $dz$.}. $W^{(0)}_{i}\left(x\right)$ has a
compact support and can be explicitly evaluated by
\begin{flalign}
&\!\!\!\!
W^{(0)}_{i}\left(x\right)= \left\{
\begin{matrix}
W^{(0)}_{i^{-}}\left(x\right)=\frac{\left(x-x_{i-1}\right)}{\Delta
x}, & \text{for~}x_{i-1}\leq x\leq x_{i}
\\
W^{(0)}_{i^{+}}\left(x\right)=\frac{\left(x_{i+1}-x\right)}{\Delta
x}, & \text{for~}x_{i}\leq x\leq x_{i+1}
\\
0, & \text{elsewhere}
\end{matrix}
\right.. \nonumber
\end{flalign}
Since unknowns at boundary grid points are related by
\begin{flalign}
\phi^{(\kappa)}_{N^{(0)}}W^{(0)}_{{N^{(0)}}^{-}}\left(x_{N^{(0)}}\right)=\phi^{(\kappa)}_{1}W^{(0)}_{1^{+}}\left(x_{1}\right)e^{i\theta_{0}},
\end{flalign}
one can implement (\ref{eq:1D_B-PBC}) by putting
$\phi^{(\kappa)}_{N^{(0)}}=\phi^{(\kappa)}_{1}$ and introducing a modified basis function
\cite{Sukumar2009classical}
\begin{flalign}
W^{(0)}_{{N^{(0)}}^{-}}\left(x\right) = e^{i\theta_{0}}
\left(x-x_{N^{(1)}}\right)/\Delta x
\end{flalign}
for $x_{N^{(1)}}\leq x\leq x_{N^{(0)}}$. Similar to the FDM case,
the total number of DoFs in the FEM formulation also becomes
$N^{(1)}$. Plugging (\ref{eq:field_FEM_rep}) into
(\ref{eq:1-D_helmholtz_wave_eq}) and applying integral by parts and Galerkin testing
\cite{strang2008analysis} over $\Omega$, one can obtain
(\ref{eq:FDM_helmtholz_mat}) where
\begin{flalign}
&\left\{\phi^{(\kappa)}\right\}_{i} =\phi^{(\kappa)}_{i}
\\
&\left[\mathcal{S}\right]_{i,j} = -\int_{\Omega}dx \left[
\frac{d}{dx}\left\{W^{(0)}_{i}\left(x\right)\right\}^{*}\frac{d}{dx}W^{(0)}_{j}\left(x\right)
\right],
\\
&\left[\mathcal{M}\right]_{i,j} = \int_{\Omega}dx \left[
\epsilon\left(x\right)
\left\{W^{(0)}_{i}\left(x\right)\right\}^{*}W^{(0)}_{j}\left(x\right)
\right], \label{eq:Galerkin_method}
\end{flalign}
for $i=1,2,\cdots,N^{(1)}$ and $j=1,2,\cdots,N^{(1)}$.
Note that the derivative-periodic condition can be confirmed from the derivation of the stiffness matrix in which the boundary integral term is set to be zero.

\section{Validity of numerical normal modes}
\subsection{Numerical orthonormality}
\begin{figure*}
\centering \subfloat[by
FDM]{\includegraphics[width=.5\linewidth]{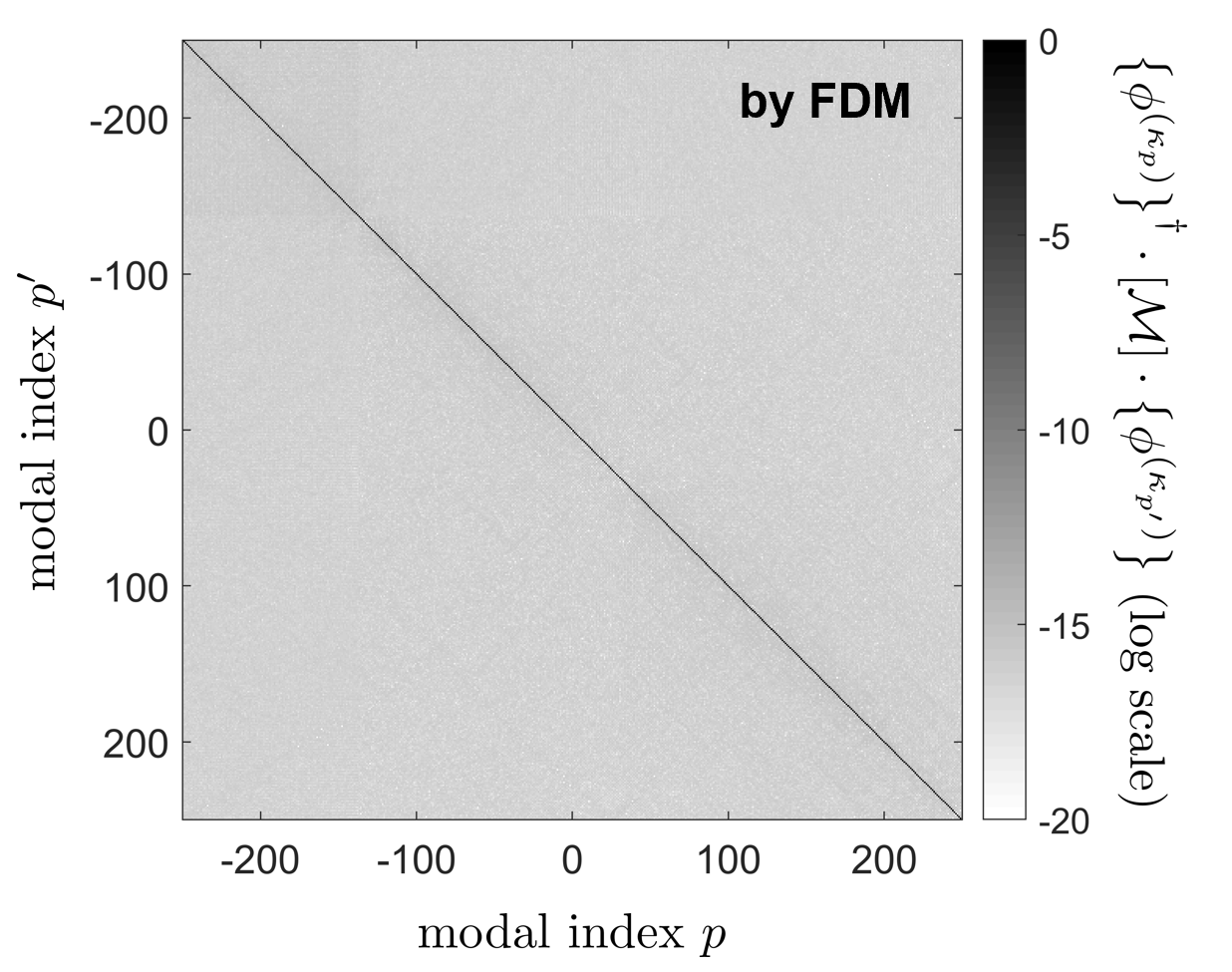}}
\subfloat[by
FEM]{\includegraphics[width=.5\linewidth]{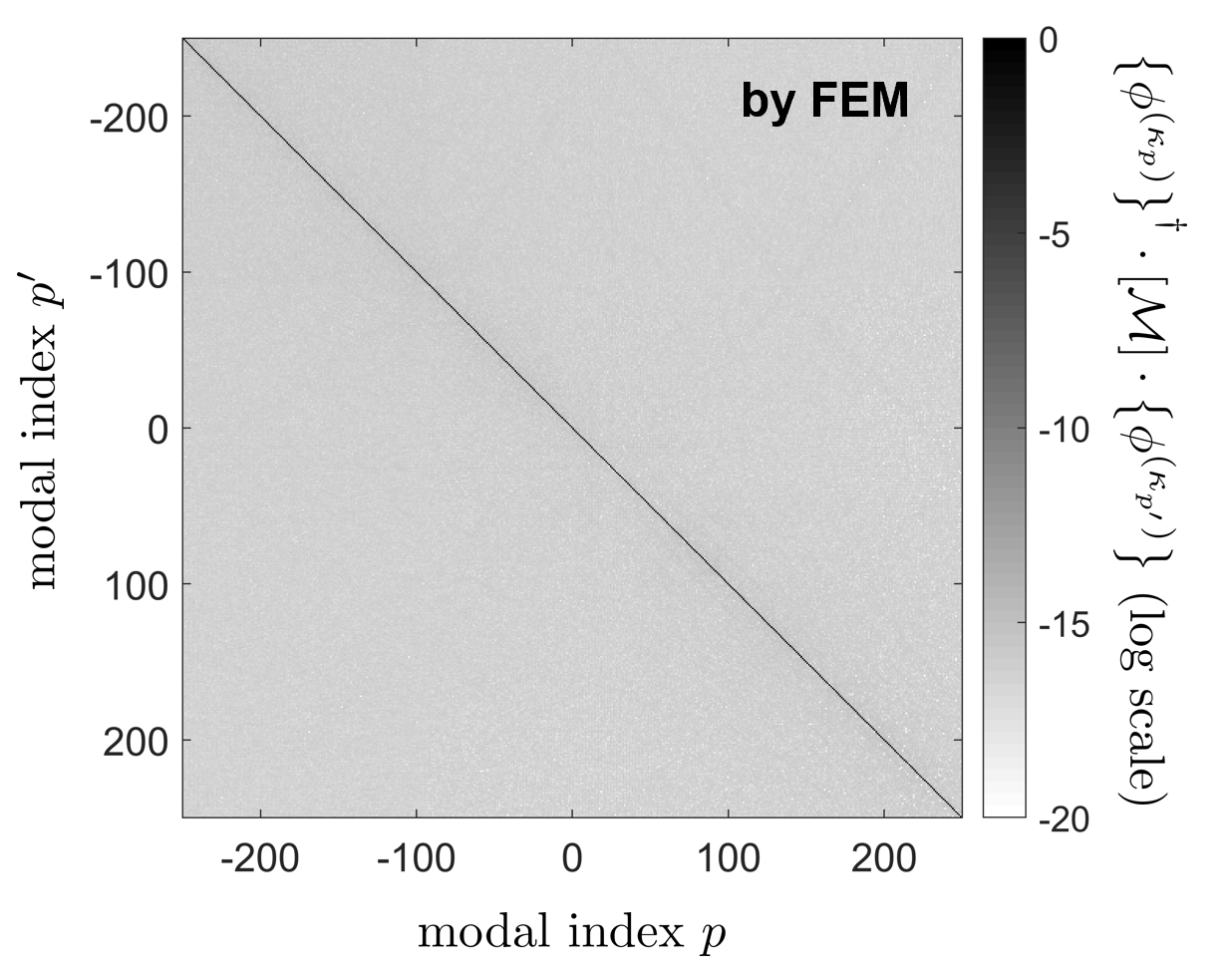}}
\caption{The orthonormal property of numerically-evaluated spatial
eigenfunctions.} \label{fig:BF_orthonormality}
\end{figure*}

Fig. \ref{fig:BF_orthonormality} shows results of
(\ref{eq:ortho_rel_1D}), measured in a decimal log scale,
obtained by FDM and FEM formulations. It can be seen that all
diagonal elements, very close to $1$, are larger than the other
elements by around $10^{15}$; thus, our numerical spatial
eigenfunctions preserves the orthonormal property to within machine
precision. Note that, in this example, $R_{x}=3$ [m], $N^{(0)}=501$,
$\Delta x=6$ [mm], $\theta_{0}=\pi/2$ [rad], $R_{s}=0.3$ [m],
$\epsilon_{s}=20$, and $N_{s}^{(0)}=51$.

\subsection{Dispersion diagram for bi-periodic medium}
\begin{figure}
\centering
\includegraphics[width=\linewidth]{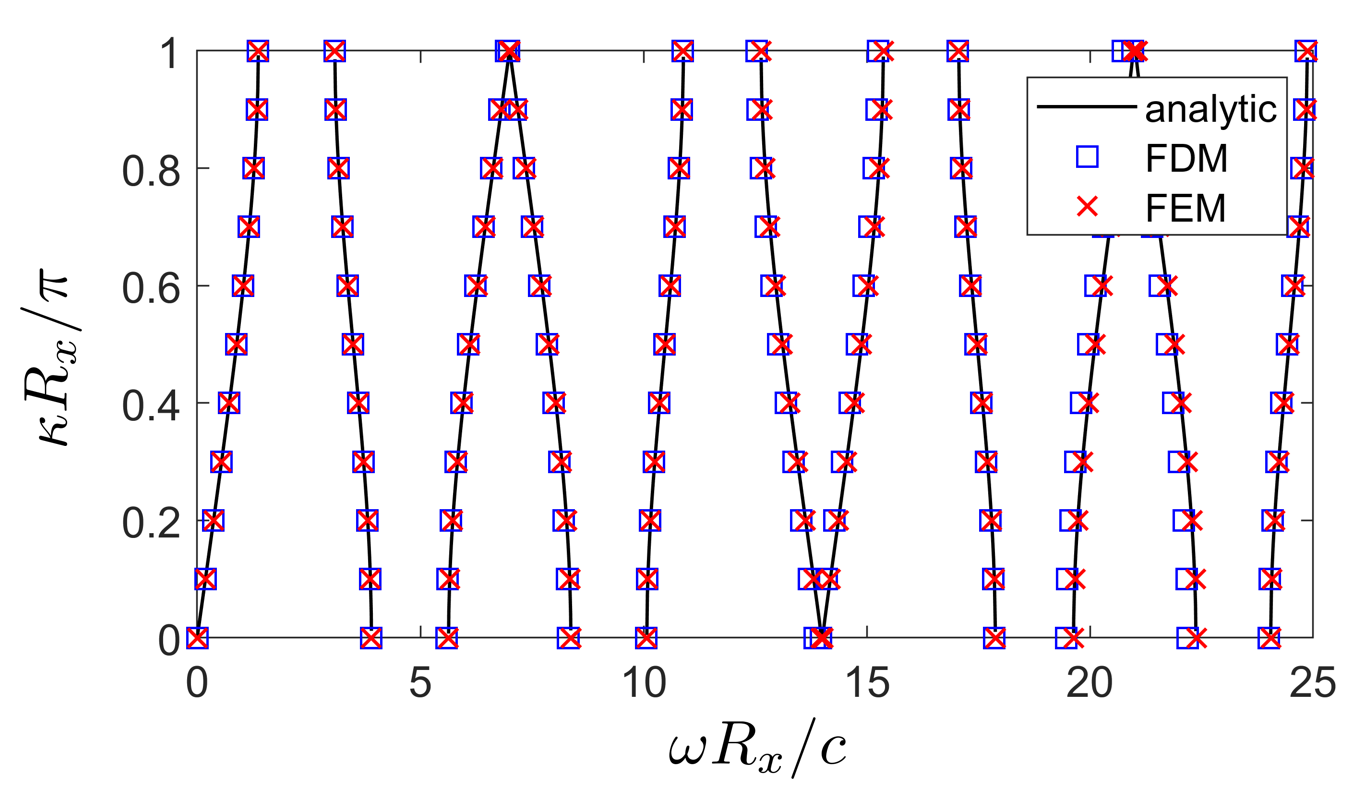}
\caption{Dispersion diagram of numerical spatial eigenfunctions.}
\label{fig:Dispersion_diagram}
\end{figure}

In order to test the validity of numerical eigenfunctions, for the
same example, we evaluate their dispersion diagrams compared with
analytic predictions. Physically speaking, the resultant dispersion
diagram is equivalent to that of Bloch-Floquet modes traveling in a
bi-periodic medium. Each eigensolver with a specific $\theta_{0}$
produces a set of solutions $\left(\kappa^{(p)},\omega_{(p)}\right)$
where $\kappa^{(p)}=\left(\theta_{0}+2\pi p\right)/R_{x}$ and
$\omega_{(p)}$. Evaluating the eigensolvers for different
$\theta_{0}$, one can finally construct the dispersion diagram by
folding all solutions within the first Brillouin zone where
$\kappa\in\left[-\pi/R_{x}, \pi/R_{x}\right]$. Fig.
\ref{fig:Dispersion_diagram} illustrates normalized dispersion
diagrams obtained by FDM, FEM, and analytic solution. Note that the
analytic solutions results from the use of transfer matrix method
(TMM). Overall, excellent agreement among them can be observed.

\section{Quantum beam splitter}
In the quasi-monochromatic limit, the composite input quantum state for the single-photon incidence can be expressed
\begin{flalign}
\ket{\psi^{(1)}}_{\text{in}}
=
\ket{1}_{1}
\otimes
\ket{0}_{0}
=
\hat{a}^{\dag}_{ {1} }\ket{\left\{0\right\}}
\label{eq:single_photon_incidence}
\end{flalign}
where subscription number denotes port index which can be regarded as
different wavenumbers in terms of the propagation direction. It is
important to note that one should account for effects of the vacuum
fluctuation in the unused port 0 even with no photon incoming. From
the Heisenberg picture, input operators (ports 1 and 0) are evolving
through a unitary operator, which incorporates scattering effects
inside the 50/50 BS.
A tensor product is often used to characterize a quantum system made up of multiple subsystems which are not interacting or not entangled.
In other words, a larger Hilbert Space for such quantum system can be constructed by the tensor product of sub-Hilbert spaces.
Then, input and output operators are associated with the Stokes relation
\begin{flalign}
\left[
\begin{matrix}
\hat{a}^{\dag}_{ {2} }
\\
\hat{a}^{\dag}_{ {3} }
\end{matrix}
\right]
=
\frac{1}{\sqrt{2}}
\left[
\begin{matrix}
1 & i \\
i & 1 \\
\end{matrix}
\right]
\cdot
\left[
\begin{matrix}
\hat{a}^{\dag}_{ {0} }
\\
\hat{a}^{\dag}_{ {1} }
\end{matrix}
\right]
\label{eq:scattering_matrx}
\end{flalign}
where the transformation is called scattering operator or matrix.
On the other hand, adopting the Schr\"{o}dinger picture, one can also find the corresponding output quantum state for a given input quantum state to the QBS.
Replacing input operators in (\ref{eq:single_photon_incidence}) by output operators as in (\ref{eq:scattering_matrx}), one can obtain the output quantum state
\begin{flalign}
\ket{\psi^{(1)}}_{\text{out}}
&=
\frac{1}{\sqrt{2}}
\left(
i\hat{a}^{\dag}_{ {2} }
+
\hat{a}^{\dag}_{ {3} }
\right)
\ket{\left\{0\right\}}
\nonumber \\
&=
i\ket{1}_{ {2} }
\otimes
\ket{0}_{ {3} }
+
\ket{0}_{ {2} }
\otimes
\ket{1}_{ {3} }.
\label{eq:single_photon_output_state}
\end{flalign}
Thus, the output quantum state is {\it entangled} since it cannot be
simply represented by the tensor product between two individual
quantum states, implying that the single photon can be present at
both output ports simultaneously before measurement. But after
measurement, the output quantum state collapses into one of two
possible states, as a result, the single photon is detected at only
one of output ports in the random sense (50\% probability), as
illustrated in Fig. \ref{fig:QBS_schm_1}.

Similarly, for a two-photon incidence at each input port, one can represent an input quantum state as
\begin{flalign}
\ket{\psi^{(2)}}_{\text{in}}
=
\ket{1}_{ {1} }
\otimes
\ket{1}_{ {0} }
=
\hat{a}^{\dag}_{ {1} }
\hat{a}^{\dag}_{ {0} }
\ket{\left\{0\right\}}
\label{eq:two_photon_incidence}
\end{flalign}
Again, viewing in the Schr\"{o}dinger picture with the use of (\ref{eq:scattering_matrx}), one can obtain the resulting output quantum state
\begin{flalign}
&\ket{\psi^{(2)}}_{\text{out}}
=
\frac{1}{\sqrt{2}}
\left(
i\hat{a}^{\dag}_{ {2} }
+
\hat{a}^{\dag}_{ {3} }
\right)
\frac{1}{\sqrt{2}}
\left(
\hat{a}^{\dag}_{ {2} }
+
i\hat{a}^{\dag}_{ {3} }
\right)
\ket{\left\{0\right\}}
\nonumber \\
&=
\frac{1}{2}
\left(
i
\hat{a}^{\dag}_{ {2} }
\hat{a}^{\dag}_{ {2} }
-
\hat{a}^{\dag}_{ {2} }
\hat{a}^{\dag}_{ {3} }
+
\hat{a}^{\dag}_{ {3} }
\hat{a}^{\dag}_{ {2} }
+
i
\hat{a}^{\dag}_{ {3} }
\hat{a}^{\dag}_{ {3} }
\right)
\ket{\left\{0\right\}}
\nonumber \\
&=
\frac{i}{\sqrt{2}}\ket{2}_{ {2} }\otimes\ket{0}_{ {3} }
+
\frac{i}{\sqrt{2}}\ket{0}_{ {2} }\otimes\ket{2}_{ {3} }.
\label{eq:two_photon_incidence_output}
\end{flalign}
It should be stressed that, in the second equality of (\ref{eq:two_photon_incidence_output}), second and third terms can be canceled out since creation operators for different modes commute.
This cancellation reflects on the wave-like nature of indistinguishable photons by interference of two possible output quantum states with different probability amplitudes, i.e., $-\frac{1}{\sqrt{2}}\ket{2}_{ {2} }\otimes\ket{0}_{ {3} }$ and $\frac{1}{\sqrt{2}}\ket{2}_{ {2} }\otimes\ket{0}_{ {3} }$, resulting from one photon is transmitted and the other is reflected.
Eventually, the output quantum state is also entangled, meaning that when two photons enter the 50/50 BS one in each input port, two photons always exit through one of output ports while being bunched (see Fig. \ref{fig:QBS_schm_2}).

One can infer more details about the behaviors of QBS from
\cite{Mandel1995optical}[Chap. 12] and
\cite{Gerry2004introductory}[Chap. 6].

%\subsection{word}
%\wordcount

\nocite{*}

\bibliography{mybibpra}% Produces the bibliography via BibTeX.

\end{document}